\documentclass[12pt]{article}
\usepackage{color}
\usepackage{graphicx}

\input xy
\xyoption{all}
\xyoption{web}

\newlength{\abstractwidth}

\renewcommand{\thefootnote}{\fnsymbol{footnote}}
\renewcommand{\thanks}[1]{\footnote{#1}}
\newcommand{\starttext}{
\setcounter{footnote}{0}
\renewcommand{\thefootnote}{\arabic{footnote}}}

\newcommand{\bea}{\begin{eqnarray}}
\newcommand{\eea}{\end{eqnarray}}
\newcommand{\ee}{\end{equation}}
\newcommand{\be}{\begin{equation}}

\newcommand{\ea}{\end{array}}
\newcommand{\bac}{\begin{array}{c}}
\newcommand{\bacc}{\begin{array}{cc}}
\newcommand{\barcl}{\begin{array}{r@{}c@{}l}}
\newcommand{\brcl}{\begin{array}{rcl}}
\newcommand{\bdm}{\begin{displaymath}}
\newcommand{\edm}{\end{displaymath}}

\newcommand{\half}{\frac{1}{2}}

\def\cO{{\cal O}}

\def\cQ{{\cal Q}}

\def\bR{{\bf R}}
\def\bZ{{\bf Z}}

\def\Re{{\rm Re}}
\def\Im{{\rm Im}}

\def\half{ {1\over 2}}
\def\p{\partial}

\def\ep{\varepsilon}

\def\l{\lambda}
\def\L{\Lambda}
\def\o{\omega}

\def\G{\Gamma}

\def\no{\nonumber}
\def\sm{\smallskip}

\def\funnel{$AdS_2$-funnel}
\def\kap{$AdS_2$-cap}

\begin{document}
\starttext
\setcounter{footnote}{0}

\begin{flushright}
IGC-11/11-2
\end{flushright}

\bigskip

\begin{center}

{\Large \bf Simple holographic duals to boundary CFTs\footnote{This work is supported in part by NSF grants
PHY-08-55356 and PHY-07-57702.} }

\vskip 0.6in

{ \bf Marco Chiodaroli$^a$, Eric D'Hoker$^b$,  Michael Gutperle$^b$}

\vskip .2in

 ${}^a$ {\sl Institute for Gravitation and the Cosmos,}\\
{\sl The Pennsylvania State University, University Park, PA 16802, USA} \\
{\tt \small mchiodar@gravity.psu.edu}

\vskip 0.2in

 ${}^b$ { \sl Department of Physics and Astronomy }\\
{\sl University of California, Los Angeles, CA 90095, USA}\\
{\tt \small dhoker@physics.ucla.edu,  gutperle@physics.ucla.edu; }

\end{center}

\vskip 0.2in

\begin{abstract}

By relaxing the regularity conditions imposed in arXiv:1107.1722 on half-BPS solutions to 
six-dimensional Type~4b supergravity, we enlarge the space of solutions to include two new 
half-BPS configurations,
which we refer to as the \kap\ and the \funnel. We give evidence that the \kap\ and \funnel\ can be 
interpreted as fully back-reacted brane solutions with respectively $AdS_2$ and $AdS_2\times S^2$ 
world volumes. \kap\ and \funnel\ solutions with a single asymptotic $AdS_3 \times S^3$ region are
constructed analytically. We argue that \kap\ solutions provide simple examples of holographic duals to 
boundary CFTs in two dimensions and present calculations of their holographic boundary entropy
to support the BCFT dual picture.

\end{abstract}

\newpage


\baselineskip=17pt
\setcounter{equation}{0}
\setcounter{footnote}{0}

\newpage

\section{Introduction}
\setcounter{equation}{0}
\label{sec1}

In a preceding paper \cite{Chiodaroli:2011nr} we gave an exact construction of the 
general local half-BPS solution to six-dimensional Type 4b supergravity, with $m$ 
tensor multiplets\footnote{As will be discussed in section \ref{sec2}, all values of 
$m$ are allowed classically,  but only the supergravities with $m=5$ and $m=21$ are anomaly free.}, 
on $AdS_2 \times S^2$ warped over a two-dimensional surface with boundary $\Sigma$. 
By imposing certain regularity and topology conditions, such as requiring the presence 
of $N$ asymptotic $AdS_3 \times S^3$ regions, we obtained a general class of 
{\sl regular half-BPS string-junction solutions}.
These solutions are holographically dual to an arrangement of $N$ two-dimensional 
CFTs, living on $N$ half-lines, which are all joined together at a single point. The regular 
string-junction solutions have a moduli 
space whose dimension $2(m+1)N -m-2$ is accounted for precisely by the number of 
three-form charges and the number of un-attracted scalars of the tensor multiplets.

\sm

The boundary of the moduli space is reached as the regularity conditions are 
stretched to their limit. Since the regularity conditions on half-BPS string-junction solutions 
include strict inequalities, the corresponding moduli space is naturally an open space, 
and the degeneration limits of regular solutions will not, in general, be regular in the 
original sense. If these limiting solutions exhibit natural mathematical and physical features,
it may become appropriate to {\sl compactify the moduli space} by including some, if not all, 
such {\sl generalized solutions}. 

\sm

The study of degenerations of regular half-BPS string-junction solutions indicates that the key 
limiting case indeed admits a natural physical interpretation. This case arises as follows. The 
three-form charge $\mu^A$ associated with one of the asymptotic $AdS_3 \times S^3$ regions is a 
vector under the $U$-duality group $SO(5,m)$ of Type 4b supergravity. For regular 
solutions we must have $\mu \cdot \mu >0$, since this quantity\footnote{Throughout, we shall denote 
the $SO(5,m)$-invariant metric by $\eta ={\rm diag} (I_5, - I_m)$ and its associated inner 
product between two $SO(5,m)$ vectors by a center dot.}  sets the scale for the
radii of the asymptotic $AdS_3$ and $S^3$ spaces, as well as for
the central charge of the dual CFT. The limit in which $\mu \to 0$ leads again to a
regular solution, but with one fewer asymptotic $AdS_3 \times S^3$ region. However, 
since the $SO(5,m)$-invariant metric has indefinite signature, it is possible to take a limit in 
which $\mu \cdot \mu \to 0$ while $\mu \not= 0$. Although the associated $AdS_3 \times S^3$ 
asymptotic region still disappears in this limit, and the central charge of the dual CFT still tends 
to zero, a non-trivial charged configuration remains, which we shall refer to as an \kap. 

\sm

The possibility of extending the moduli space of regular half-BPS string junction solutions by 
including \kap s raises the question as to whether other natural extensions of moduli space
may be of mathematical and 
physical relevance. A second natural extension indeed exists: it is obtained by allowing for 
half-BPS solutions with asymptotic regions which are locally isometric to 
$AdS_2 \times S^2 \times S^1 \times \bR^+$. In view of  the cylinder-like geometry of 
the $S^1 \times \bR^+$ factor of this space-time, we shall refer to this configuration 
as an \funnel.

\sm

Both the \kap\ and the \funnel\ solutions may be characterized in terms of the holomorphic one-form 
$\Lambda^A$ of the half-BPS string junction solutions of
\cite{Chiodaroli:2011nr}. The \kap\ corresponds to an extra pole on the boundary of $\Sigma$
with three-form charge vector $\mu$ satisfying $\mu \cdot \mu=0$, 
while  the \funnel\ corresponds to an extra pole in the interior of $\Sigma$, with 
three-form charge vector satisfying $\mu \cdot \mu >0$.

\sm

In this paper, the charges and space-time metric for the simplest {\sl generalized solutions}
including these new configurations will be derived. The corresponding geometries have 
a single asymptotic $AdS_3 \times S^3$ region, and have either two \kap s or one \funnel.  
We will confirm that an \kap\ gives a finite contribution to the entanglement entropy 
which matches exactly the result from a weak-coupling computation in the dual BCFT. 

\sm

In the remainder of this introduction, we shall present arguments involving superalgebras
and supersymmetric probe-branes in an $AdS_3 \times S^3$ space-time, 
to support the dual BCFT picture for the \kap\ and a plausible setting for the \funnel.

\subsection{Superalgebras and probe-branes}

The global symmetry of the $AdS_3 \times S^3$ vacuum solution of Type 4b supergravity 
is the superalgebra $PSU(1,1|2) \times PSU(1,1|2)$, which is inherited as (part of) the asymptotic symmetry 
of each asymptotic $AdS_3 \times S^3$ region 
\cite{Maldacena:1997re,VanProeyen:1986me,D'Hoker:2008ix}. Of course, the full asymptotic
symmetry is enlarged to the Brown-Henneaux  Virasoro algebras \cite{Brown:1986nw}, as expected for a dual 
to a two-dimensional CFT. The global symmetry of the regular half-BPS string-junction 
solutions, as well as of the generalized solutions developed in this paper, is reduced to 
a single copy of $PSU(1,1|2)$, whose maximal bosonic subalgebra $SO(1,2) \times SO(3)$ 
is the isometry algebra of $AdS_2 \times S^2$, and whose number of fermionic generators accounts
for 8 supersymmetries. The conformal 
algebra $SO(1,2)$ is indeed suitable for a dual interface or BCFT interpretation,
as it is the symmetry of a linear boundary or interface in two dimensions, while the 
$SO(3)$ factor corresponds to the associated reduced R-symmetry.

\sm

One of the simplest ways to reduce the symmetries is to consider probe-branes 
in the $AdS_3\times S^3$ background (recall that 
in the probe-brane approximation 
the back-reaction of the probe onto the supergravity background is not being taken into 
account). The reduced bosonic symmetry indicates that 
a probe-brane should have either an $AdS_2$ world volume, where the location  
of the probe on $S^3$ breaks the $SO(4)$ isometry to $SO(3)$, or an 
$AdS_2 \times S^2$ world volume inside $AdS_3\times S^3$.
A  comprehensive  analysis of supersymmetric probe $D$-branes preserving 
8 of 16 supersymmetries in a  $AdS_{3}\times S^3 \times K3$ background of 
Type IIB supergravity 
was carried out in \cite{Raeymaekers:2006np} (see also 
\cite{Bachas:2000fr,Raeymaekers:2007vc}) . The results are summarized 
in the left four columns of Table 1 below. Here, the internal manifold $M_4$ is ether $K3$ or $T^4$,
and $C_2$ denotes a supersymmetric two-cycle inside $K3$ onto which the 
probe-brane is wrapped. 
\begin{table}[ht]
\begin{center}
\begin{tabular}{|c||c|c|c||c|c|}
\hline
brane&$AdS_{3}$&$S_{3}$& $M_4$ & charges & supergravity solution\\
\hline \hline
D1&$ AdS_{2}$ &$\cdot$&$\cdot$ & $\mu \cdot \mu = 0 $ & \kap\ \\
\hline
D5&$ AdS_{2}$ &$\cdot$&$M_4$ & $\mu \cdot \mu = 0 $ & \kap\  \\
\hline
D3&$ AdS_{2}$ &$\cdot $&$C_{2}$ & $\mu \cdot \mu = 0 $ & \kap\  \\
\hline
D3&$ AdS_{2}$ &$S_{2}$&$\cdot$ & $\mu \cdot \mu \neq 0 $ & \funnel\  \\
\hline
D7&$ AdS_{2}$ &$S_{2}$&$M_4$ & $\mu \cdot \mu \neq 0 $ & \funnel\ \\
\hline
\end{tabular}
\caption{Correspondence between half-BPS probe-brane 
configurations and fully back-reacted generalized half-BPS string junction 
supergravity solutions.} 
\label{table2}
\end{center}
\end{table}

One goal of the present paper is to construct  
fully back-reacted solutions corresponding to these probe-branes. In the 
last column of Table 1, we have indicated the {\sl generalized 
half-BPS string junction solution} which we propose to associate with each 
probe-brane configuration.

\subsection{Holographic duals to BCFTs and to interface CFTs}

The holographic dual to a boundary CFT is closely related to the holographic
dual to an interface CFT. Janus solutions \cite{hep-th/0304129}  to supergravity provide concrete dual
realizations of interface CFTs. Specifically, the supersymmetric Janus solution
dual to a two-dimensional CFT has precisely the same symmetries and geometrical
structure as the one expected from a BCFT in the same dimension: 
$AdS_2 \times S^2$ warped over a Riemann surface $\Sigma$ with boundary.
In terms of local coordinates $x, y$ on $\Sigma$, the metric  is given by, 
\bea
\label{1b1}
ds^{2}= f_{1}^{2 } ds^{2}_{AdS_{2}} + f^{2}_{2}ds^{2}_{S^{2}} +\rho^2 (dx^2+ dy^2)
\eea
The vacuum solution $AdS_3\times S^3$ corresponds to a flat strip $\Sigma = \{ x+iy, \, 
x \in \bR, \, y \in [0,\pi]\}$, and metric factors given by $f_1^2= \cosh^2x$, 
$f_2^2 = \sin^2 y$, and $\rho^2=1$. At the boundary of $\Sigma$, given by $y=0,\pi$, 
the volume of the sphere $S^2$ vanishes. In the regions $x\to \pm \infty$, 
the $AdS_2$ metric factor blows up and an asymptotic $AdS_3$ is formed. 

\sm

On the one hand, the Janus solution corresponds to a deformation of the 
vacuum solution in which various supergravity fields can take different asymptotic values in
the two  $AdS_3\times S^3$ regions. As such, it gives a concrete realization for 
the holographic dual to a conformal interface where two CFTs living on half-spaces are glued 
together along a line. 

\sm

On the other hand, a BCFT is a two-dimensional CFT on a half-space which 
terminates along a one-dimensional boundary.  Recently a proposal for the holographic 
realization of a BCFT was made in \cite{Takayanagi:2011zk,Fujita:2011fp}, building 
upon the original proposal given in \cite{Karch:2000gx}. In its simplest realization the 
holographic BCFT proposal utilizes a Janus-like $AdS_2$ slicing of $AdS_3$ as in (\ref{1b1}).   In \cite{Takayanagi:2011zk,Fujita:2011fp}, the space-time is cut off in the bulk of the space,  
so that the second  $AdS_3 \times S^3$ asymptotic region of the Janus solution is effectively removed. 
Consequently, only one boundary component is retained, thereby producing a geometry 
suitable for a dual BCFT on a half-space.   
The location of the extra boundary to the bulk space is associated with the presence 
of  branes (and possibly orientifold planes  \cite{Fujita:2011fp}).  
In \cite{Takayanagi:2011zk,Fujita:2011fp}, the proposal was used to 
calculate BCFT observables such as correlation functions and  the boundary 
entropy (or $g$-function). One key result of the present paper is to provide  
a new, and  simple,  string-theoretic realization of the proposal of 
\cite{Takayanagi:2011zk,Fujita:2011fp},
in terms of the \kap\ generalizing the half-BPS string junction solutions 
found in \cite{Chiodaroli:2011nr}.

\subsection{Organization}

The remainder of this paper is organized as follows. In Section \ref{sec2}, we provide a brief
review of six-dimensional Type 4b supergravity, and of the local and global
half-BPS string-junction solutions of \cite{Chiodaroli:2011nr}. We also obtain 
explicit formulas for the holographic entanglement entropy of the string junctions. 
In Section \ref{sec3}, we motivate and spell out the precise relaxed regularity 
conditions for the generalized solutions. The new geometrical objects will be
referred to as the \kap\ and the \funnel.
The local geometry and charges of the \kap\ and the \funnel\ are derived as well.
In Sections \ref{sec4}, an  explicit solution is produced for the case of  a single
asymptotic $AdS_3 \times S^3$ region  with two  \kap s.
The holographic entanglement entropy is evaluated, it is compared to the 
boundary entropy of the BCFT, and used to further support the proposed 
correspondence.  In Section \ref{sec6},  the solution with a single $AdS_3 \times S^3$ 
and an arbitrary number of \kap s is constructed explicitly using a light-cone parametrization
in terms of auxiliary poles.   In Section \ref{sec5b}, an explicit solution for  a single
asymptotic $AdS_3 \times S^3$ region with one \funnel\ is derived, and the 
corresponding holographic entanglement entropy is calculated.
Concluding remarks are given in Section \ref{sec7}. Some calculational details are relegated to several appendices.

\newpage

\section{Half-BPS string-junction solutions}
\setcounter{equation}{0}
\label{sec2}

In this section we shall review the salient features of the six-dimensional Type 4b, or (0,4),
supergravity, and of the regular string-junction solutions constructed in 
\cite{Chiodaroli:2011nr}. We shall also produce a simple formula for the holographic 
entanglement entropy of these solutions.

\subsection{Six dimensional Type 4b supergravity}\label{sec22}

The supersymmetry generators of the $(0,4)$ theory consist of 4 complex self-conjugate 
Weyl spinors of the same chirality, which may be organized into 2 symplectic-Majorana 
multiplets. The theory was constructed in \cite{Romans:1986er}, and here we shall follow the
conventions of \cite{Chiodaroli:2011nr}.    
The field contents of the $(0,4)$ theory consists a supergravity multiplet, and $m$ tensor 
multiplets\footnote{Throughout, the index $A=(I,R)$ will label the fundamental 
representation of $SO(5,m)$, while the indices $(i,r)$ will label the fundamental representation 
of $SO(5) \times SO(m)$. Their ranges are given by, $I,i=1,\cdots , 5$ and $R,r=6, \cdots, m+5$ respectively.}
The supergravity multiplet contains the metric and the Rarita-Schwinger field as well as  
rank-two anti-symmetric tensors $B^I$. Each tensor multiplet contains an anti-symmetric rank-two tensor 
$B^R$, and a quartet of Weyl fermions whose chirality is opposite to that of the gravitini.
Finally, the scalar fields  live in an $SO(5,m)/SO(5)\times SO(m)$ coset and are
parameterized by a frame field $V^{i,r}_{\;\;A}$. The field strengths associated with the 
anti-symmetric tensor fields obey the following 
self-duality relations, 
\bea
H^i = V^i_{\; A} \; dB^A & \hskip 1in & \star H^i = + H^i
\no \\
H^r = V^r_{\; A}\; dB^A && \star H^r = - H^r
\eea
Classically, all values of $m$ are allowed, but anomaly cancellation restricts $m$ to the 
values  5 or 21. Both cases may be obtained by compactifying Type IIB supergravity, the $m=5$ 
case on $T^4$; the $m=21$ case on $K3$.  The theory has a Minkowski vacuum where all 
anti-symmetric tensor fields are set to zero and one can construct self-dual BPS string solutions 
which preserve eight of the 16 Minkowski supersymmetries.  In the near horizon limit the 
$AdS_3\times S^3$ vacuum emerges with 16 supersymmetries and $PSU(1,1|2)^2$ symmetry superalgebra.

\subsection{The general local half-BPS solutions}\label{sec23}

As argued in \cite{Chiodaroli:2010ur}, a junction of $N$ dyonic strings 
in six flat dimensions can preserve four supersymmetries provided that the strings are oriented according 
to the $SO(5)$ component of their $SO(5,m)$ charge vectors and that a tension balance condition is obeyed.  
The near-horizon geometries of these junctions are given by half-BPS string-junction solutions 
which are invariant under eight residual supersymmetries.

\sm

The general local half-BPS solution, 
with $SO(1,2) \times SO(3)$ symmetry on the space $AdS_2 \times S^2 $ warped 
over a two-dimensional Riemann surface $\Sigma$ with boundary, was obtained in \cite{Chiodaroli:2011nr}. 
The metric and anti-symmetric tensor fields of the solution take the invariant form, 
\bea
\label{ansatz}
ds^2 & = & f_1^2 ds^2 _{AdS_2} + f_2 ^2 ds^2 _{S^2} +  ds^2 _\Sigma 
\no \\
B^A & = & \Psi ^A \, \o_{AdS_2} + \Phi ^A \, \omega _{S^2}
\eea
Here, $ds^2 _{AdS_2}$ and $ds^2 _{S^2}$ are the invariant metrics respectively on the spaces 
$AdS_2$ and $S^2$ of unit radius, while $\o_{AdS_2}$ and $\omega _{S^2}$ are the corresponding 
volume forms. In local complex coordinates $(w,\bar w)$ on $\Sigma$, the metric  
$ds_\Sigma ^2 = \rho ^2 |dw|^2$ is parametrized by a real function~$\rho$. 

\sm

The general local solution is specified completely in terms of a real positive harmonic function 
$H$ on $\Sigma$, and $m+2$ meromorphic functions $\lambda^A$ on $\Sigma$.  Equivalently,
one may use $m+2$ holomorphic one-forms $\Lambda ^A$, which are related  to $\lambda ^A$ by, 
\bea
\Lambda ^A = \lambda ^A \, \p H
\eea 
It will often be convenient to go back and forth between the use of $\l^A$ and $\L^A$.
As discussed in \cite{Chiodaroli:2011nr},  the structure of the BPS equations and the reality of the fields impose the 
restriction $\Lambda^3=\Lambda^4=\Lambda^5=0$, up to $SO(5)$ rotations, 
as well as the following constraints, 
\bea
\label{1b2}
\l \cdot \l & = & 2
\no \\
\bar \l \cdot \l & \geq & 2
\eea
The scalar fields take values in an $SO(2,m)/SO(2)\times SO(m)$ sub-manifold of the 
full  scalar coset space; the anti-symmetric tensor fields  are restricted accordingly,
$B^3=B^4=B^5=0$.
The solution of \cite{Chiodaroli:2011nr} then provides explicit formulas for the metric factors,
\bea
\label{7c1}
f_1 ^4 & = & H^2 ~ 
{ \bar \lambda \cdot \lambda  +2 \over \bar \lambda \cdot  \lambda  -2} 
\no \\
f_2 ^4 & = & H^2 ~ 
{ \bar \lambda \cdot  \lambda  -2 \over \bar \lambda \cdot \lambda  +2} 
\no \\
\rho^4 & = & {|\partial _w H|^4 \over 16 H^2} 
(\bar \lambda \cdot  \lambda  +2)  ( \bar \lambda \cdot \lambda  -2)
\eea
The solutions for the real-valued flux potential functions $\Phi ^A$ and $\Psi ^A$ are as follows,
\bea
\label{7c3}
\Phi ^A =  -   \sqrt{2} \, {  H \, \Re( \lambda^A) 
\over \bar \lambda \cdot \lambda +2 } + \tilde \Phi ^A 
& \hskip 0.6in &
\tilde \Phi^A = { 1 \over 2 \sqrt{2} } \int \Lambda^A   +{\rm c.c.}
\no \\
\Psi ^A =  - \sqrt{2} \,   { H \, \Im (\lambda^A) 
\over \bar \lambda \cdot \lambda -2 } + \tilde \Psi ^A 
& \hskip 0.6in &
\tilde \Psi^A = { i \over 2 \sqrt{2} } \int  \Lambda^A  +{\rm c.c.}
\eea
In this paper we will not utilize the expressions for the scalars, which may be found in 
\cite{Chiodaroli:2011nr}.

\subsection{Regularity and topology conditions}\label{sec24}

The regularity requirements imposed in \cite{Chiodaroli:2011nr} 
are reflected on the data $H,\l^A$ as follows, 
\begin{description}
\itemsep -0.03in
\item{~~~1.} In the interior of $\Sigma$ we have $H>0$ and $\bar \lambda \cdot \l >2$;
\item{~~~2.} On the boundary $\p \Sigma$ of $\Sigma$ we have $H=0$ and $\Im (\lambda ^A) =0$,
except at isolated points;
\item{~~~3.} The one-forms $\Lambda ^A$ are holomorphic and nowhere vanishing in the interior 
of $\Sigma$, forcing the poles of $\lambda^A$ to coincide with the zeros of $\p_w H$;
\item{~~~4.}  The functions $\l^A$ are holomorphic near $\p \Sigma$, allowing $\Lambda ^A$ 
to have poles on $\p \Sigma$ only at those points where $\p_w H$ has a pole.
\end{description}
The conditions on $H$ restrict its poles $x_n$ to be located on the real axis, to be simple, 
and to have positive residues $c_n$. Solutions with $N$ poles are parametrized as follows, 
\bea
\label{1b3}
H(w,\bar w) =  \sum _{n=1}^N  \left ({ i \, c_n \over w - x_n} - { i \, c_n \over \bar w - x_n} \right )
\hskip 1.1in c_n >0
\eea
The conditions on $\l^A$ and $\Lambda ^A$ dictate the following form for $\Lambda ^A$, 
\bea 
\label{covariant} 
\L ^A (w) =  \sum _{n=1}^N \left ( { -i \kappa^A _n \over (w-x_n)^2} +  { -i \mu^A _n \over w-x_n} \right ) 
\hskip 1in 
\sum _{n=1}^N \mu_n ^A=0
\eea
where the residues $\kappa ^A_n$ and $\mu_n^A$ are real.\footnote{Following the notation of 
\cite{Chiodaroli:2011nr}, we shall suppress the basic differential factor $dw$ in writing 
$\Lambda ^A$ throughout.} Equations (\ref{1b2}) impose non-linear 
equalities between $x_n, c_n, \kappa_n^A, \mu_n^A$, and restrict their range by inequalities. They 
may be solved explicitly using a parametrization in terms of auxiliary poles, as 
will be summarized in Section \ref{sec6}.

\sm

The space-time geometry in the neighborhood of a pole $x_n$ may be read off by setting
$w= x_n + r e^{i \theta}$ and considering the leading $ r \to 0$ asymptotics. The metric behaves as follows, 
\bea
\label{8c3}
ds^2 \sim  \sqrt{ 2 \mu_n \cdot \mu _n } \left (  {dr^2 \over r^2} 
 + {2 c_n^2 \over \mu_n \cdot \mu_n } {1\over r^2} ds_{AdS_2}^2+ d\theta^2 + \sin^2\theta \, ds_{S^2}^2 
 \right )+ \cO (r^2)
\eea
which confirms that the space is asymptotically isometric to $AdS_3 \times S^3$.
The radii of these spaces is set by $ \mu_n \cdot \mu _n $, requiring this quantity to be positive 
for any regular solution. The three-form Page charge, carried by the $AdS_3$ throat
at the pole $x_n$, is given by,
\bea
\label{7g1}
Q^A _n  
\equiv  \oint _{S^3} dB^A= i \sqrt{2} \pi   \int _{x_n} \!\! \L^A + {\rm c.c.}
=  2 \sqrt{2} \pi^2   \mu ^A _n
\eea
The central charge of the dual CFT carried by the corresponding $AdS_3$ is given by,
\bea 
c = {3 \, Q_n \cdot Q_n \over 16 \pi^2 G_N }  = {3 \pi^2 \, \mu_n \cdot \mu_n \over 2 G_N }   
\label{central} 
\eea
a quantity which will be useful in the sequel.

\subsection{Holographic entanglement entropy}

The holographic entanglement entropy for a regular string-junction solution may be defined in terms 
of the area of a minimal surface in the bulk geometry which subtends a partition of the system 
on the boundary. This definition follows a proposal set forth is \cite{Ryu:2006bv}.\footnote{The 
entropy for an interface Janus solution and for string-junction solutions with scalar 
fields restricted to a $SO(2,2) / SO(2) \times SO(2)$ sub-manifold of the scalar coset, 
were derived respectively in \cite{Chiodaroli:2010ur} and \cite{Chiodaroli:2010mv}.}
The minimal surface of interest to us here is supported on a single point in the $AdS_2$ factor.
As a result, it sits at a fixed value of the holographic parameter, at fixed time, and is
everywhere space-like. The holographic entanglement entropy is then given  by the area of the minimal surface spanned by $\Sigma\times S^2$ which evaluates to,
\bea
\label{7c0}
S_\ep  = { 1 \over 4 G_N} \int _{\Sigma _\ep} |dw|^2 \rho^2 \int _{S^2} f_2^2 \, \hat e^{23}
\eea
Here, $G_N$ is Newton's constant in six dimensions. The subscript $\ep$ is used here to
indicate that the above entropy integral requires regularization, to be spelled out shortly.
Using the results of (\ref{7c1}) for the metric functions, 
we obtain a natural $SO(5,m)$-invariant expression, 
\bea
\label{7c2}
S _\ep = {\pi \over  G_N} \int _{\Sigma _\ep} |dw|^2 \, |\p_w H |^2  \left ( \l \cdot \bar \l -2 \right )
\eea
Close to the asymptotic region labeled by $n$, the entropy density on $\Sigma$ takes the from,
\bea
\label{7d1}
{ \pi \over G_N} \rho^2 f_2^2 |dw|^2 \sim \pi \, { \mu_n \cdot \mu_n \over 2 G_N}  \, 
{ ( \Im w )^2 |dw|^2 \over |w-x_n|^4}
\eea
which diverges in the neighborhood of $w=x_n$. To regularize the entropy integral
while maintaining minimality of the bulk geometry surface $\Sigma _\ep \times S^2$, we 
choose to cut off $\Sigma$ along a geodesic of the effective metric $ \rho^2 f_2^2 |dw|^2$,
and denote the regularized surface by $\Sigma _\ep$.

\sm

To evaluate the $\ep$-dependence of the resulting entropy integral, we change local coordinates from $w$
to $0<r$ and $0 \leq \theta < \pi$, with $w = x_n + r e^{i \theta}$. One can show that, in the limit of small $r$,
the geodesic equation admits the solution $r=$\, constant. As a result, the surface $\Sigma _\ep$ is
obtained from $\Sigma$ by removing half-disks of coordinate radius $r=\ep$ around the poles $x_n$.
Using this construction, we derive the $\ep$-dependence of the entropy integral, and find, 
\bea
\label{7d4}
S_\ep \sim - { \pi^2 \over 4 G_N} (\mu_n \cdot \mu_n) \ln \ep
+ \cO(\ep ^0)
\eea
Note that the regularization near $x_n$ depends only on $\ep$ and the {\sl local charge data} $\mu_n$.
 
\newpage

\section{Relaxing the regularity conditions}
\setcounter{equation}{0}
\label{sec3}

The regular half-BPS string-junction solutions reviewed in the previous section were 
constructed so that the supergravity fields are manifestly regular everywhere 
on $\Sigma$. In particular, the regularity of the one-forms $\Lambda ^A$ inside 
$\Sigma$ and on the boundary $\p \Sigma$ (except at isolated points on $\p \Sigma$)
guarantees that all three-form charges have support only on the asymptotic regions 
$AdS_3 \times S^3$. These regular solutions correspond to fully
back-reacted junctions of dyonic strings in six dimensions. 

\sm

However, half-BPS solutions based on the  Ansatz (\ref{ansatz}) 
should be expected to exist also for stacks of pure $Dp$-branes (with $p=1,3,5$ or $7$) 
in the curved $AdS_3 \times S^3 \times K3$ space-time. As summarized in Table 1, 
these branes can be arranged in half-BPS configurations which preserve the isometries of 
the $SO(2,1)\times SO(3)$-invariant Ansatz of (\ref{ansatz}). 
In contrast to the regular solutions corresponding to dyonic strings
of the preceding section, solutions corresponding to pure $Dp$-branes should have 
only one kind of brane charge. They are expected to have regular supergravity fields,
except for the known mild singularities coinciding with the world-volume of the branes. 

\sm

In the next subsection, we shall characterize the three-form charge vectors $\mu$
associated with pure $Dp$-branes, and argue that they must satisfy $\mu \cdot \mu=0$
while $\mu \not= 0$. Asymptotic regions with $\mu \cdot \mu =0$ were excluded from
the regular solutions of Section \ref{sec2}, but they can be naturally included by relaxing 
the regularity conditions imposed in Section \ref{sec24}. We shall spell out the
relaxed regularity conditions for such generalized solutions in Section \ref{sec32} below. 
Equivalently, these generalized solutions may be obtained 
as degenerations of regular solutions, and thus play an important mathematical
role in the compactification of the moduli space of regular solutions.
We shall argue that, physically, the generalized solutions provide full string theoretic 
holographic BCFTs in two dimensions.

\subsection{Physical meaning of $\mu \cdot \mu$}

In our construction of a BCFT, a key role is being played by the charge vector $\mu$,
and in particular by its ``square" $\mu \cdot \mu$. For example, as already announced in Table 
\ref{table2} of the introduction, the \kap\ configuration is characterized by $\mu \cdot \mu =0$.

\sm

To clarify the physical meaning of $\mu \cdot \mu$, and in particular of 
the special case $\mu \cdot \mu =0$, it will be advantageous to look 
at the behavior of branes in the full Type IIB superstring theory,
compactified on  $M_4=K3$ or $T^4$. Its effective low energy theory is the 
six-dimensional Type 4b supergravity with $m$ tensor multiplets. As explained in 
\cite{Dijkgraaf:1998gf,Seiberg:1999xz}, in string theory, 
the scalar moduli space is given by the coset, 
\bea
SO(5,m,\bZ) \backslash SO(5,m, \bR) / SO(5)\times SO(m) 
\eea
where $m=5$ for $M_4=T^4$ and $m=21$ for $M_4=K3$. 
Here, $K=SO(5,m,\bZ)$ is the corresponding $U$-duality group. 
The charges $v$ of the three-form field take values in $ \bR^{5,m}$ and
transform as vectors under $SO(5,m,\bR)$. We denote the associated
invariant inner product of two charge vectors $v$ and $w$ by $v \cdot w$, as before.
The allowed charges for a six-dimensional string must lie in an 
even self-dual lattice $\Gamma^{5,m}$ so that $v \cdot v \in 2 \bZ$.
The inner product naturally splits lattice vectors $v$ into purely  "space-like" and 
purely ``time-like" components,  
\bea
v = v_+ + v_-  \hskip 1in v_+ \in \bR^{5,0} \qquad v_- \in \bR^{0,m}
\eea
The condition for a six-dimensional string to be BPS reduces to, 
\bea
v \cdot v = 2 N 
\eea
where $N$ is a non-negative integer.
The tension of the string is given by $T={\rm const}\;  |v_+|$.

\sm

An important example is provided by the $D1/D5$-brane system. 
In this case, the lattice $\G^{5,m}$ is given by a direct sum
of two lattices with indefinite signature,   
\be
\Gamma^{5,m} = \Gamma^{1,1} \oplus \Gamma^{4,m-1}
\ee
The lattice $\Gamma^{4,m-1}$ is fixed, but the lattice $\Gamma^{1,1}$ 
parameterizes the $D1$- and $D5$-brane charges, 
respectively denoted by $Q_1$ and $Q_5$, and we have,
\be
N=Q_1Q_5
\ee
The near-horizon limit of the $D1/D5$ system is described by a two-dimensional 
${\cal N}=(4,4)$ super-conformal field theory with central charge $c=6 N = 6 Q_1 Q_5$.
When one kind of charge vanishes, namely either $Q_1=0$ or $Q_5=0$, 
the corresponding charge vector is null, $v \cdot v =0$. This is a key observation
for us, which may be framed in more general terms as follows.

\sm

All vectors $v$  for a given value of $v \cdot v$ can be mapped into one another other 
by a $U$-duality transformation.  Since the BPS condition already requires $v \cdot v \geq 0$,
there are only two distinct classes of BPS strings in six-dimensions:
\begin{enumerate}
\item Strings having null charge vector, i.e. $v \cdot v = 0$. 
These strings can be mapped to a $D1$-brane by a $U$-duality transformation.
\item Strings having space-like  charge vector , i.e. $v \cdot v > 0$. These strings are in the 
$U$-duality orbit of a  $D1/D5$ bound state with given charges $2 Q_1 Q_5 = v \cdot v$.   
\end{enumerate}

The three-form charge vectors $\mu$ arising in our string-junction solutions
are proportional to the lattice charges $v$, but the quantization is not seen
at the supergravity level. The condition for a (regular) BPS solution is $\mu \cdot \mu >0$.
Based on the above arguments, the physical nature of an asymptotic region 
characterized by charge vector $\mu$ may be clarified from the value taken by $\mu \cdot \mu$.
The lesson to be drawn is that $\mu \cdot \mu=0$ corresponds to an object 
which is $U$-dual to a $D1$-brane. In the case of probe-branes in an $AdS_3 \times S^3$ background, 
the first three entries of Table 1 belong to this $U$-duality orbit with $\mu \cdot \mu =0$.

\sm

A different analysis is required for branes wrapping a two-sphere,
which corresponds to the last two entries in Table \ref{table2}. 
As argued in \cite{Bachas:2001vj,Raeymaekers:2007vc}, these $AdS_2 \times S^2$ branes 
have the same charges as a dyonic string, and hence are characterized by $\mu \cdot \mu > 0$.

\subsection{Allowed singularity types for generalized solutions}\label{sec32}

In this section, we shall spell out the detailed relaxed regularity conditions which will allow
for the inclusion of both the \kap\ and the \funnel\ solutions,
\begin{description}
\itemsep -0.03in 
\item{~~~1.} In the interior of $\Sigma$ the data satisfy $H>0$ and $\bar \lambda \cdot \l >2$;
\item{~~~2.} On the boundary $\p \Sigma$ of $\Sigma$ one has $H=0$ and $\Im (\lambda ^A) =0$,
except at isolated points;
\item{~~~3'.} The one-forms $\Lambda ^A$ are meromorphic and nowhere vanishing in the interior 
of $\Sigma$, forcing  $\lambda^A$ to have a pole wherever  $\p_w H$ has a zero;
\item{~~~4'.}  The functions $\l^A$ are meromorphic near $\p \Sigma$, as well as in the interior of $\Sigma$.
\end{description}
Conditions 1. and 2. are identical to those listed for the regular solutions in
Section \ref{sec24}, while conditions 3'. and 4'. relax conditions 3. and 4. of Section \ref{sec24}.
The half-BPS solutions satisfying the above conditions will be referred to as {\sl generalized solutions}.
They  will be allowed to exhibit two new classes of singularities in $\Lambda^A$ or $\lambda^A$:  
\begin{description}
 \item{ ~~~ \underline{The \kap} }\\
is produced by a pole in  $\lambda^A$  on $\p \Sigma$. 
We shall show that the $SO(5,m)$ charge vector $\mu$, given by the residues at this pole, satisfies $\mu \cdot \mu = 0$,
thus producing a vanishing extensive contribution to the entanglement entropy. Each pole corresponds to the fully back-reacted
solution of a probe-brane  in the $U$-duality orbit of a fundamental strings, and has $AdS_2$ world-volume.   
 \item{~~~\underline{The \funnel\  }} \\
is produced by a pole in $\lambda^A$ at a point in the interior of $\Sigma$ where $\p_w H$ is nonzero. We shall show that 
the residue of the pole satisfies $\mu \cdot \mu \neq 0$, and that the corresponding solution has a
$AdS_2 \times S^2 \times S^1 \times \mathbf{R}^+ $ funnel corresponding to the back-reaction of 
$D3$-branes with $AdS_2 \times S^2$ world-volume.
\end{description}
In the particular case of an $AdS_3 \times S^3$ background, generalized solutions 
are sufficient to account for all the possible probe-branes listed in Table \ref{table2}.
However, the \kap\ and \funnel\ solutions  can also be added to any one of the regular solutions
summarized in Section \ref{sec2}. The above generalized solutions include a natural compactification 
of the moduli space of regular solutions.

\subsection{The \kap}
\label{sec33}

For the \kap, the extra pole in $\l^A$ is located at a point $u$ on the boundary $\p \Sigma$.
In the neighborhood of $u$, the meromorphic function $\l^A$ behaves as follows, 
\bea 
\lambda^A(w) \sim {p^A \over w- u} + q^A - s^A (w-u)   \hskip 1in p^A \not= 0
\eea
The condition $\Im (\lambda ^A)=0$ on $\p \Sigma$ requires the coefficients 
$p^A, q^A, s^A$ to be real, while the condition $\l \cdot \l = 2$ imposes the relations
$p \cdot p  = p \cdot q = 0$ and $  q \cdot q = 2 + 2 \ p \cdot s$.
Finally, the condition $\bar \lambda \cdot \lambda >2$ in the interior of $\Sigma$ is 
satisfied provided we  have $p \cdot s >0$.

\sm

Alternatively, the \kap\ solution may be obtained from a regular solution for which 
$u$ coincides with a pole $x_n$ common to  $H$ and $\Lambda$, 
and then taking the limit in which, 
\bea 
c_n  = \kappa_n^A = 0 \hskip 1in  A = 1,2,6 \dots m+5 
\eea
but keeping $\mu^A_n \neq 0$ at that pole.  
The condition $\lambda \cdot \lambda =2$ then requires,
\bea
\mu_n \cdot \mu_n = 0 
\label{charges-type1} 
\eea
The local behavior of the relevant six-dimensional metric
factors may be easily exhibited using the change of variables $w=u + r e^{ i \phi}$,
and we find,
\bea 
f_1^4 &=& 4 (p \cdot s) { \big | \p_w H (u)  \big |^{2} } 
\left( 1 + (p \cdot s) \sin^2 \phi   \right) r^2  
\no \\
f_2^4 &=& 4 (p \cdot s) {  \big| \p_w H (u) \big|^{-2}  \sin^4 \phi \ r^2 \over 
1 + (p \cdot s)  \sin^2 \phi} 
\no \\
\rho^4 &=& { 1 \over 4} ( p \cdot s ) \big| \p_w H (u) \big|^2 
\left( 1 + (p \cdot s) \  \sin^2 \phi \right) { 1 \over  r^2 }  
\eea
This solution has a curvature singularity at a finite geodesic distance from any 
point in the interior of $\Sigma$.
The fact that the charges satisfy $\mu_n \cdot \mu_n =0$, 
but are otherwise generic, suggests that we should interpret the singularity 
as produced by a probe-brane with $AdS_2$ world-volume which is in the $U$-duality 
orbit of a fundamental string  (or $D1$-brane). 
Singularities of this sort cover the first four entries of Table \ref{table2}. 

\sm

This interpretation can be confirmed by uplifting the solutions to ten-dimensions. 
The relation between the ten-dimensional and six-dimensional string-junction 
solutions constructed in \cite{Chiodaroli:2009yw} and \cite{Chiodaroli:2011nr} 
respectively, is known in the particular case in which the scalar fields live 
in a $SO(2,2)/SO(2)\times SO(2)$ coset. Fortunately, the leading (singular) 
behavior of fields and metric factors close to an \kap\ can always be mapped to 
the one of a $SO(2,2)/SO(2)\times SO(2)$ solution using a $U$-duality transformation. 
We shall study the expansion of the fields close to an \kap \ singularity in Appendix 
\ref{appenda}, and find some solutions where the fields have exactly the radial 
dependence expected for a fundamental $D1$- or $D5$-brane.       

\sm

Finally, the general solution with \kap s may be constructed 
in the auxiliary pole parametrization; this will be carried out in Section \ref{sec6}.

\subsection{The \funnel}

For the \funnel\ solution, the extra pole in $\l^A$ is located at a point $u$ in the interior 
of~$\Sigma$, with $\Im \ u > 0$, at which we have $\p_w H(u) \not=0$. In the neighborhood 
of $u$, the meromorphic functions $\lambda ^A$ behave as, 
\bea\label{funnelpole}
\lambda^A  (w) \sim { p^A \over w- u} + q ^A - s^A (w-u) 
\hskip 1in p^A \not= 0
\eea
The expansion coefficients $p^A, q^A, s^A$ are allowed to take complex values, 
and must satisfy $p \cdot p  = p \cdot q = 0$ and $ q \cdot q = 2 + 2 \ p \cdot s$
in view of the relation $\l \cdot \l =2$. Furthermore,  the condition $\bar \l \cdot \l >2$ 
in the interior of $\Sigma$ will be satisfied in a neighborhood of $u$ when $\bar p \cdot p > 0$, or when 
$\bar p \cdot p= \bar p \cdot q=0$ and $\bar q \cdot q - | \bar s \cdot p + \bar p \cdot s | >2$.

\sm

The pole (\ref{funnelpole})  produces $AdS_2$ and $S^2$  factors with finite and equal 
radii at $w=u$,  which follows from $f_1^2 = f_2^2 = H(u) \not=0$.
When $\bar p \cdot p >0$,  the pole  generates a semi-infinite funnel on $\Sigma$,
governed by the metric,
\bea
ds_\Sigma ^2 = \rho^2 |dw|^2 \sim \rho_u^2 \left | { dw \over w - u} \right |^2
\hskip 1in 
\rho^2_u = \bar p \cdot  p \, { | \p_w H (u) |^2 \over H (u) } >0
\label{metric-case2-2} 
\eea
Its geometry is seen perhaps more clearly via the change of variables 
$w=u + e^{ - \zeta +i \theta}$ with $\zeta >0 $ and $0 \leq \theta < 2 \pi$, 
in terms of which we have, 
\bea
 ds_\Sigma ^2 \sim  \rho_u^2 (  d \zeta ^2  + d \theta ^2  )
\eea
Thus, as anticipated, the local geometry of the solution near w=u has the form
\bea
AdS_2 \times S^2 \times S^1 \times \bR^+
\eea
and may be viewed as a degeneration of the asymptotic $AdS_3 \times S^3 $ region.
Note that this metric is regular at $w=u$, however this point is an infinite geodesic 
distance from other points in $\Sigma$.  Some properties of the behavior
of the funnel will explored further in Section \ref{sec5b}, where an explicit global 
solution will be presented.

\sm

The pole at $w=u$ in $\lambda^A(w)$ produces a three-form charge density vector 
$\cQ^A$,  which has support on the $AdS_2$ and $S_2$. It is given by integrating 
the three-form field along a closed curve surrounding the pole $u$, 
\bea
\cQ^A  & = & \cQ^A_{AdS_2} \, \o_{AdS_2} + \cQ^A _{S^2} \, \o_{S^2} 
\eea
Where the charge densities can be expressed as follows \cite{Chiodaroli:2011nr},
\bea
\cQ_{AdS_2}^A & = & - {1 \over \sqrt{2}} \Im \oint _{u} \L^A = \sqrt{2} \pi \Re \left ( p^A \p H(u) \right )
\no \\
\cQ_{S^2}^A & = & {1 \over \sqrt {2}} \, \Re  \oint _{u} \L^A  = \sqrt{2} \pi \Im \left ( p^A \p H(u) \right )
\eea
Integrating the charge density $Q_{S^2}^A$ over the unit two sphere $S^2$ gives 
$4 \pi \cQ_{S^2}^A$. This quantity is the total magnetic charge $Q^A$ of the funnel, familiar from
the regular solutions which carry such charges in their $AdS_3 \times S^3$ 
asymptotic regions. The quantity $\cQ_{AdS_2}^A$ is new, however, and did not occur 
in the regular solutions. It corresponds to an electric charge density along the non-compact $AdS_2$.

\newpage

\section{Holographic realization of a BCFT}
\setcounter{equation}{0}
\label{sec4}

For a {\sl regular solution}, each pole $x_n$ of the harmonic function $H$ is associated with 
an asymptotic $AdS_3\times S^3$ region, and each three-form charge $\mu_n^A$ is carried by 
an asymptotic region, labeled by $n=1,\cdots, N$. Charge conservation in the full solution implies that, 
\bea
\label{4c1}
\sum_{n=1}^N \mu_n^A=0
\eea
It is therefore impossible to construct a holographic dual, with a single $AdS_3 \times S^3$ region,
to a boundary CFT from {\sl regular solutions}. Indeed, any non-vacuum regular solution must have 
at least two asymptotic $AdS_3 \times S^3$ regions, and is therefore dual to an interface or a 
string-junction CFT instead. However, generalized solutions with $N=1$ do exist.  
As discussed in Section \ref{sec33} an \kap\ has nonzero charge  vector $\mu^A$ which is null. 
It is possible to obey charge conservation by adding two \kap s whose null charges add up to 
the charge of a single $AdS_3 \times S^3$ region. Consequently,  by relaxing the regularity conditions, 
we are able to construct a simple holographic realization of a BCFT with only one boundary and $N=1$.

\sm

\begin{figure}[htb]
\begin{center}
\includegraphics[width=5in]{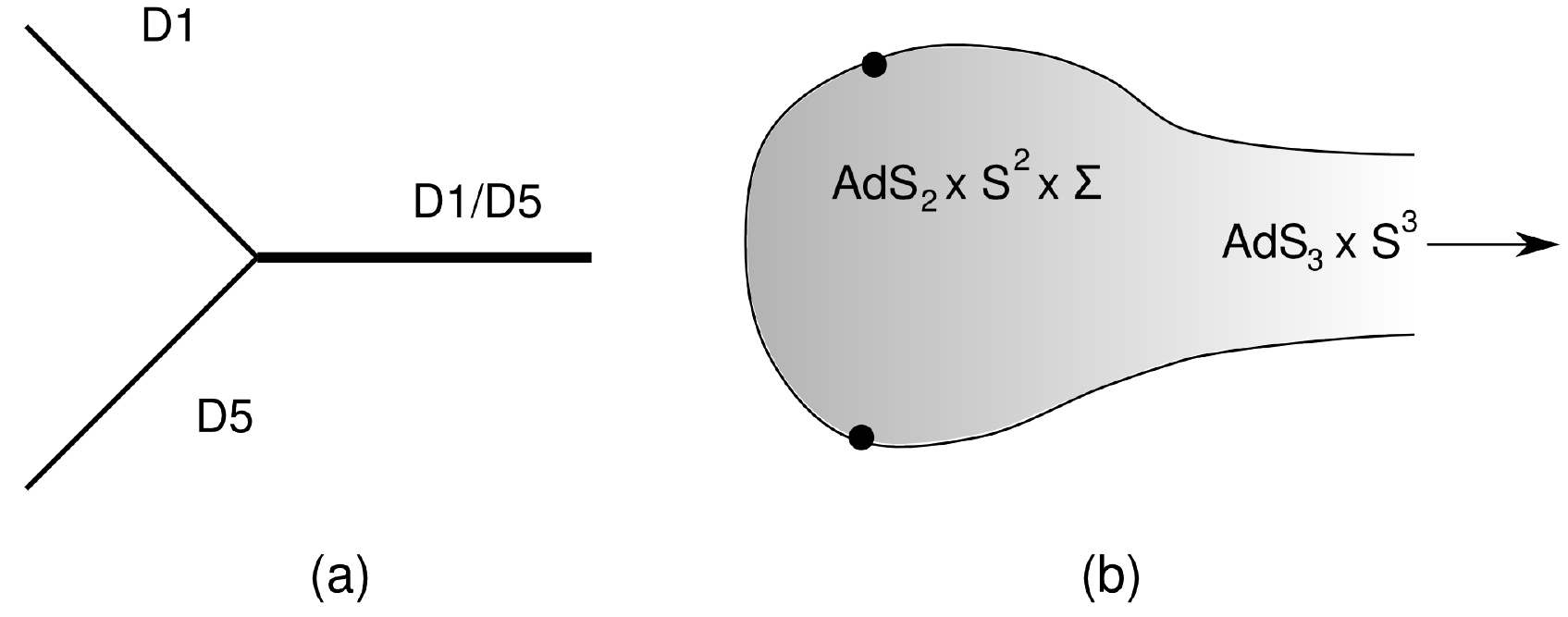} 
\caption{(a) Junction of a $D1$-brane and a $D5$-brane  forming a bound state; 
(b) The corresponding supergravity solution with two $AdS_2$-caps and one asymptotic $AdS_3 \times S^3$.} 
\label{fig1}
\end{center}
\end{figure}

From a ten-dimensional perspective, the simplest example of this kind of solution 
would give the back-reaction of a junction \cite{Dasgupta:1997pu,Sen:1997xi}, where a 
$D1$-brane and a $D5$-brane wrapping the $K3$ manifold or the four-torus come together 
to form a $D1/D5$ bound state. This is illustrated in Figure \ref{fig1}. The near-horizon 
geometry of the bound state would produce the $AdS_3 \times S^3$ asymptotic region, 
while the two branes would correspond to the two \kap s. 

\subsection{The \kap\ solution with one asymptotic $AdS_3 \times S^3$}\label{sec41}

For the geometry to have only one asymptotic $AdS_3 \times S^3$ region, 
the harmonic function $H$ should have only one pole on the real axis. 
As argued above the simplest solution obeying charge conservation has two additional caps. 
By $SL(2,R)$ symmetry of the upper half-plane, we can fix the location of the pole of $H$ at $w=0$, and  
the  locations  of the caps at $w=\pm 1$.  

\sm

The  analysis of the previous section readily implies that the residues $\kappa ^A_{\pm 1} $ 
of $ \Lambda ^A$ at the two extra poles must vanish, and that $\mu_{\pm1} \cdot \mu_{\pm1}=0$.  
We consider the following Ansatz for the harmonic function $H$ and holomorphic forms $\Lambda ^A$,
\bea
\label{4a1}
H&=& i {c_0 \over w} - i { c_0 \over \bar w} 
\no \\
\Lambda^A &=& 
- i {\kappa_{0}^A\over w^2} -i {\mu_{0}^A\over w} -i {\mu_{1}^A\over w-1}  -i {\mu_{-1}^A\over w+1}
\eea
where $c_0, \kappa _0 ^A, \mu _0 ^A$, and $\mu_{\pm 1}$ are real, and $c_0>0$.
Charge conservation implies,
\be
\label{4a2}
 \mu_0^A + \mu_1^A + \mu _{-1}^A=0
\ee
The first condition of (\ref{1b2}) imposes the following constraints,
\bea
\label{4a3}
 \kappa_0 \cdot  \kappa_0 &=& 2 c_0^2
\no \\
\kappa_0 \cdot \mu_{0}&=&0
\no \\
\mu_1 \cdot \mu_1 = \mu_{-1} \cdot  \mu_{-1} &=& 0
\no \\
2 \kappa_0  \cdot (\mu_{1} - \mu _{-1} )  &=&  \mu_0  \cdot \mu_{0}
\eea
The second condition of (\ref{1b2})  will be satisfied provided 
\bea
\label{4a4}
\mu_0 \cdot \mu_{0}>0
\eea
Conditions (\ref{4a2}), (\ref{4a3}), and (\ref{4a4}) may be solved in terms of a 
subspace of vectors which transform under the subgroup $SO(2,2)$, for example 
by setting the components with index $A=8, 9, \cdots m+5$ equal to zero.
The solutions may be exhibited in terms of two real parameters 
$\kappa>0$ and $\mu$, and we find, 
\bea
c_0 & = & \kappa / \sqrt{2}
\no \\
\kappa_0&=&\big( \kappa ,0,0,0 \big) 
\no \\
\mu_0 &=& \big( 0, 2 \kappa \mu ,0,0 \big) 
\no \\
\mu_{\pm 1}&=& \Big( \pm \kappa \mu^2  , - \kappa \mu  , \mp \kappa \mu  \sqrt{1+ \mu^2},0 \Big) 
\eea
The metric functions are determined as follows,
\bea
f_1^4 &=&   {2 \kappa ^2 \over \mu_0^2} 
{ \mu_0^2 \Im (w)^2+  \kappa ^2 |1-w^2|^2 \over |w|^4}
\no \\
f_2^4 &=&   {2 \kappa ^2 \mu_0^2 \, \Im (w)^4 \over |w|^4 
\big(\mu_0^2 \, \Im (w)^2+ \kappa^2 |1-w^2|^2 \big)}
\no \\
\rho^4&=& {\mu_0^2\over 8 \kappa ^2} { \mu_0^2 \, \Im (w)^2
+ \kappa ^2 |1-w^2|^2 \over |w|^4 |1-w^2|^4}
\eea
where we have used the abbreviation $\mu_0^2 = \mu_0 \cdot \mu_0$ throughout. 
\begin{figure}[t]
\begin{center}
\includegraphics[width=6.5in]{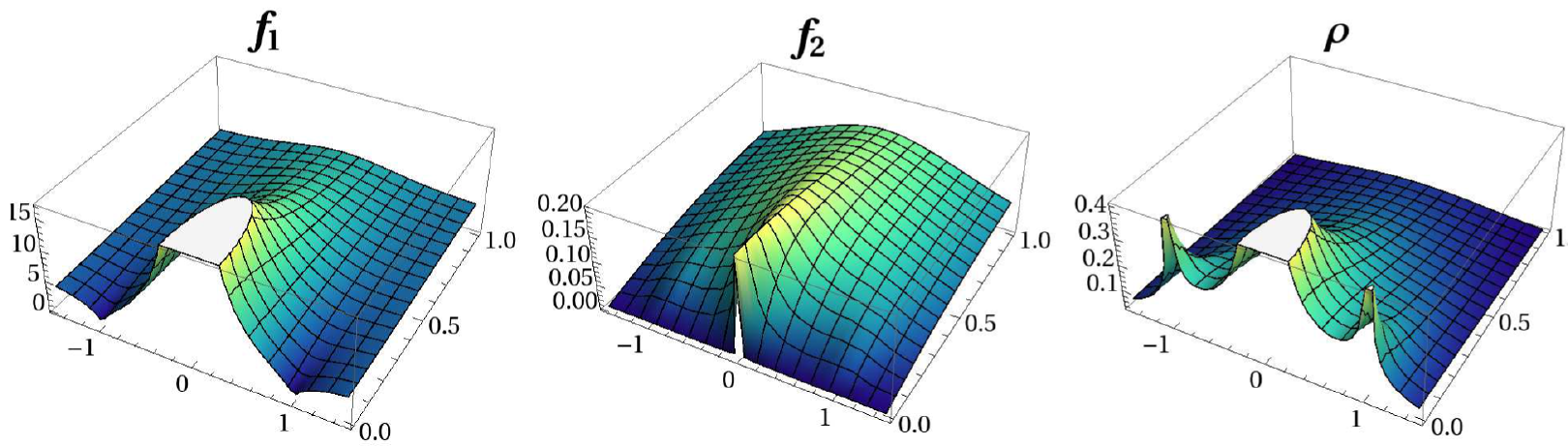} 
\caption{Plot of the metric factors for a solution with two \kap\ singularities at $w=\pm 1$ and
one asymptotic region at $w=0$. The parameters have the random values $\kappa = 0.74369$  and $ \mu = 0.020045 $.} 
\label{plotcap}
\end{center}
\end{figure}
\sm
We display plots of the metric factors for random values of the parameters in Figure \ref{plotcap}.

\subsection{Calculation of the boundary entropy}

\begin{figure}[tb]
\begin{center}
\includegraphics[width=5.2in]{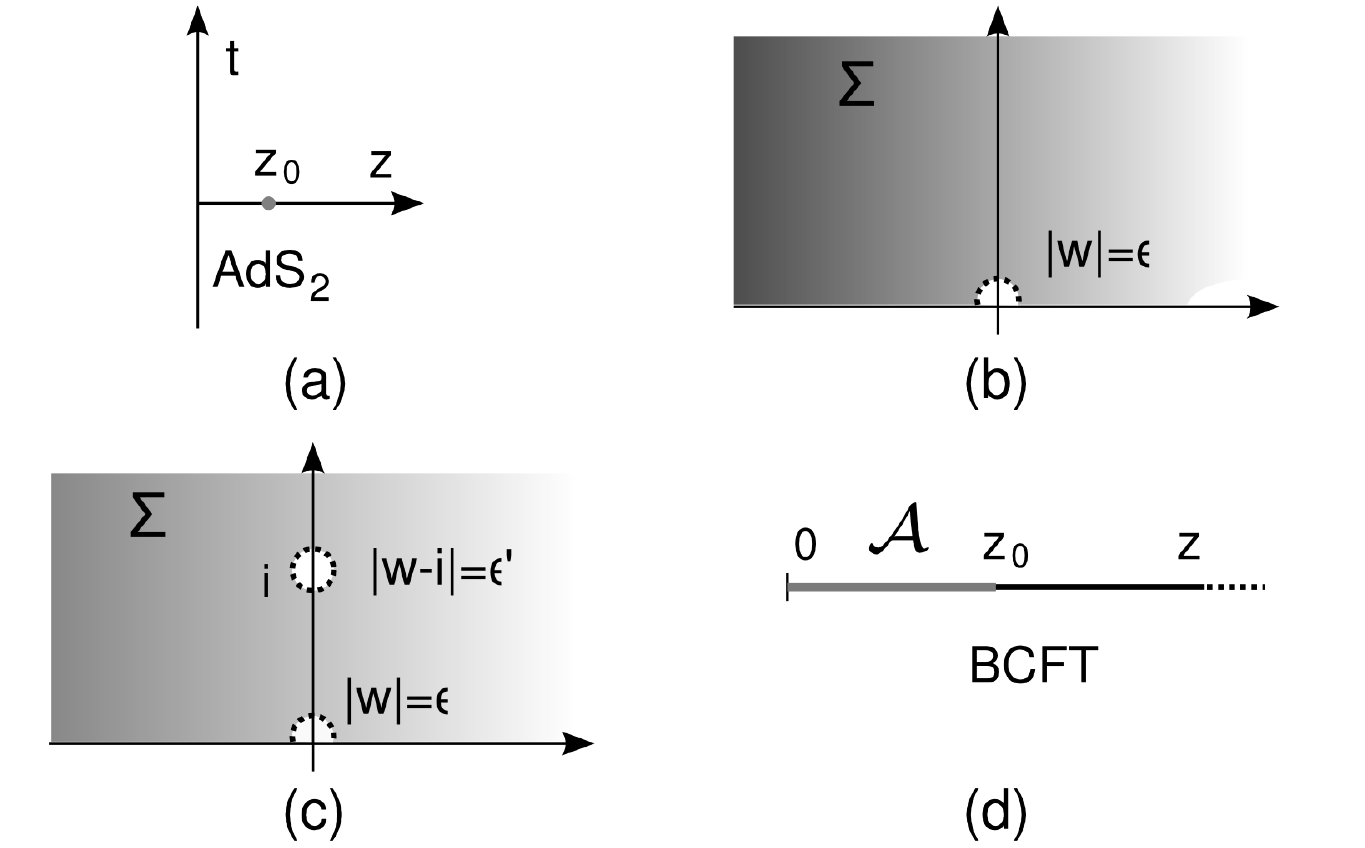} 
\caption{Minimal surface for the holographic entanglement entropy computation; 
(a) The minimal surface is a point in the $AdS_2$ space; 
(b) The minimal surface for the $AdS_2$-cap with corresponding cutoff; 
(c) The minimal surface for the 
$AdS_2$-funnel with corresponding cutoffs;
(d) Corresponding partition of the BCFT.} 
\label{fig2bcft}
\end{center}
\end{figure}

We use (\ref{7c2}) to obtain a simple expression for the holographic 
entanglement entropy, 
\bea 
S = {  \pi \mu_0^2 \over 2 G_N} 
\int_{\Sigma_\epsilon} { |dw|^2 \ \Im (w)^2 \over |w|^4 |1-w^2|^2} 
\eea
The integral is logarithmically divergent at $w=0$ and we have to introduce a cutoff  $|w|> \ep$, 
as illustrated in Figure \ref{fig2bcft}. Note that the integration  converges at the \kap\ points $w=\pm 1$.  
The holographic entanglement entropy 
can be found with the help of the following regularized integrals, 
\bea
\int _{\Sigma _\ep} |dw|^2 { 1 \over |w|^2} & = & { \pi \over 2} \ln { R^2 \over \epsilon^2}
\no \\
\int _{\Sigma _\ep} |dw|^2 { 1 \over (w-x)(\bar w -y)} & = & { \pi \over 2} \ln {  R^2 \over (x-y)^2}
\no \\
\int _{\Sigma _\ep} |dw|^2 { 1 \over (w-x)(\bar w -y)^2 } & = & { \pi \over x-y} 
\label{integrals} \eea
where $R$ is a large $|w|$ cutoff, $\ep$ is a UV cutoff, and $x,y \in \bR$. 
We obtain the following expression,
\bea \label{boundenta}
S =  { \pi^2 \mu_0^2 \over 4 G_N} \left ( \ln {1 \over \ep} - \ln 2 + \half \right )   
\eea

As discussed in \cite{Azeyanagi:2007qj,Chiodaroli:2010ur} the cutoff in the bulk has to be related 
to the UV in the CFT as follows. After a change of variables $w=e^{x+ i \phi}$,   the $AdS_3 \times S^3$ asymptotic 
region  $w=0$  is mapped into $x\to \infty$ and the metric  (\ref{8c3}) behaves as,
\bea
\label{asymmet2}
ds^2 = \sqrt{2 \mu_0^2} \left ( dx^2 + d\phi^2 +\sin^2\phi \, ds_{S^2}^2+  e^{2x}   
{\kappa_0^2\over \mu^2} {dz^2-dt^2\over z^2} \right ) + \cO (e^{-x})
\eea
The boundary of the CFT is located at $z=0$ and the region for which we calculate  
the entanglement entropy is the interval $z\in [0,z_0]$. 
As explained in Appendix  \ref{appendb},  mapping the $AdS$-slicing coordinates to 
Poincar\'e slicing  coordinates allows to relate the cutoff $\ep=e^{-x_\infty}$ to the CFT cutoff 
$\xi_{UV}$ as follows,
\be
{1\over \xi_{UV} } = {\kappa_0\over z_0 \mu_0} {1\over \ep} 
\label{relcutoffs}\ee
This relation  introduces the dependence of the entanglement entropy  on the length 
of the interval $z_0$ as well as dependence on the parameter $\kappa_0$.  
\be
S=  { \pi^2 \mu_0^2 \over 4 G_N} \left( \ln {z_0 \over \xi_{UV} }+ \ln{\kappa_0 \over 2 \mu_0} +{1\over 2}\right)
\ee
To isolate the boundary entropy, we consider the difference of the entanglement entropy 
for  arbitrary value of $\kappa_0$ and a reference value $\kappa_0=1$, while keeping $\mu_0$ 
(i.e. the cosmological constant and hence the central charge), $ z_0$  (i.e. the length of the interval) 
and the UV cutoff $\xi_{UV}$  fixed.
\bea
\label{boundent1}
S_{bound} = S(\kappa_0^2)-S(\kappa_0^2=1)=  { \pi^2 \mu_0^2 \over 4
 G_N}  \ln \sqrt{ \kappa_0^2}
\eea
Note that $\kappa_0^A$ parameterizes the un-attracted scalars moduli in the asymptotic $AdS_3 \times S^3$ 
region with charges $\mu^A$.  

\subsection{BCFT interpretation}

The result for the holographic boundary entropy should be compared to a boundary entropy of the dual CFT. 
While a complete analysis will not be performed here, following  \cite{Azeyanagi:2007qj,Chiodaroli:2010ur},  the result (\ref{boundent1}) 
 can be compared with 
a  weak coupling expression for the boundary 
entropy of a BCFT with $N_B$ compact bosons of radius $R$ obeying Dirichlet 
boundary conditions \cite{Elitzur:1998va},
\bea\label{boundent2}
S_{bound}=N_B \ln {1 \over \sqrt{2 R }}
\eea
Note that in a supersymmetric setting, the number of compact bosons is related to the central charge, and in our case we have,
\bea 
N_B = {2 \over 3} c = {\pi^2 \mu_0^2 \over G_N}
\eea
Comparing the holographic result (\ref{boundent1}) and the CFT result (\ref{boundent2}) the central charge dependence match and we can identify radius dependence in (\ref{boundent2}) with the dependence on the value of the un-attracted scalar moduli $\kappa_0^2$.   For the case of the D1/D5 the relevant CFT is the $(T^4)^N /S_N$ orbifold and in \cite{Azeyanagi:2007qj,Chiodaroli:2010ur} a precise identification of the boson radius $R$ in  (\ref{boundent2}) and the asymptotic value of the dilation which takes the role of $\kappa_0^2$ in  (\ref{boundent1}) was made. Hence, the holographic boundary entropy (\ref{boundent1}) is a U-dual generalization of this result.

\newpage

\section{Explicit solutions for one $AdS_3 \times S^3$ and $N$ caps}
\setcounter{equation}{0}
\label{sec6}

In this section, we shall adapt the method of parametrization by auxiliary poles, 
constructed in  \cite{Chiodaroli:2011nr} for regular solutions, to the case of 
{\sl generalized solutions}. We shall work out explicitly the solutions with a single
$AdS_3 \times S^3$ region and an arbitrary number $N$ of \kap s.

\subsection{Auxiliary pole parametrization of regular solutions}

We begin by reviewing the solution, given in \cite{Chiodaroli:2011nr}, 
of the constraints (\ref{1b2}) in terms of auxiliary poles, and a parametrization of the 
functions $\lambda ^A$  in terms of light-cone variables $L^A$, 
\bea
\label{8a1}
\l^A = { \sqrt{2} \, L^A \over L^6} & \hskip 1in & 
\l^2 = { 1 \over L^6} \left ( + \half - L^1 L^1 + L^R L^S  \delta _{RS} \right )
\no \\ &&
\l^6 = { 1 \over L^6} \left ( - \half - L^1 L^1 + L^R L^S  \delta _{RS} \right )
\eea
where $A=1,7,8,\cdots, m+5$, and $R,S=6,7,8, \cdots, m+5$. Since the functions $\lambda ^A$ are
meromorphic, the functions $L^A$ must also be meromorphic. Moreover, the harmonic function 
$h^A = \Im (L^A)$ associated with $L^A$ must obey Dirichlet vanishing conditions, 
\bea
\label{8a2}
h^A  = 0 \quad \hbox{on} \quad \p \Sigma 
\eea
The meromorphic functions $L^A$ may be parametrized by a finite number $P$ of simple poles,
\bea
\label{8a4}
L^A(w) = \ell ^A _\infty + \sum^P _{p=1}  { \ell ^A _p \over w-y_p}
\eea
The poles $y_p$, the residues $\ell ^A_p$, and the asymptotic parameters $\ell ^A _\infty$ 
must be real. 
The above parametrization will apply to the regularity conditions of Section \ref{sec24} for regular solutions, 
as well as to the regularity conditions of Section \ref{sec32} for {\sl generalized solutions}.

\sm

For regular solutions, avoiding curvature singularities requires $L^6$ and $\p_w H$ to 
have common zeros. Therefore, we take $L^6$ to have the form,
\be 
\label{8a5}
L^6 (w) = i \ell^6_\infty {\prod_{n}^{N} (w-x_n)^2 \over \prod_{p}^P (w-y_p) } \p_w H  (w)
\ee
In turn, to avoid an extra auxiliary pole or an extra zero at infinity, we need the number of auxiliary 
poles to be related to the number of physical poles,  $P=2N-2$.
Finally, regularity of the $\l^A$ at the auxiliary poles, and constancy of the sign of $h_1<0$, 
requires the relation,
\be 
\label{8a6}
\ell^1_p =  \sqrt{ \ell_p^R \ell_p^S \delta _{RS} } 
\ee
With this parameterization, and this choice of sign, all regularity conditions of Section \ref{sec24} for 
{\sl regular solutions} will be satisfied. 

\subsection{Solutions with $N$ \kap s}

Before we adapt the light-cone auxiliary pole parametrization to the case of solutions 
with a single $AdS_3 \times S^3$ asymptotic region and $N$ \kap s, we shall present 
here the general covariant form of $\l^A$ and $\Lambda ^A$ for such solutions.
With only a single $AdS_3 \times S^3$ region, we have one double pole
in both $\p_w H$ and $\Lambda^A$, which we place at $w=x_0=0$. We shall look for 
solutions which in addition have $N$ \kap s, which correspond to no poles at all in $H$,
and to simple poles $x_n \not= 0$ in $\Lambda ^A$, 
\bea
\label{cov}
i \p_w H = { c_0 \over w^2} 
\hskip 1in 
i \Lambda ^A = { \kappa _0 ^A \over w^2} + {\mu _0 ^A \over w} 
+ \sum _{n=1}^N { \mu_n^A \over w-x_n}
\eea
As a result, we have 
\bea
\label{8a7}
\lambda ^A = {  \Lambda ^A \over \p_w H} = {w^2 \over c_0} \left ( { \kappa _0 ^A \over w^2} 
+ {\mu _0 ^A \over w} + \sum _{n=1}^N { \mu_n^A \over w-x_n} \right )
\eea
The \kap\ configuration requires $x_n$ to be real. Reality of $\lambda ^A$ for real $w$ requires 
$\kappa _0^A, \mu_0^A, \mu_n^A$ to be real as well. Charge conservation requires
\bea
\label{8a8}
\mu_0 ^A + \sum _{n=1}^N \mu_n ^A=0
\eea
The constraint $\l \cdot \l =2$ and $\bar \l \cdot \l >2$ inside $\Sigma$ must now be enforced
on these data. It is not known how to do this explicitly in the covariant formulation, but an explicit 
solution in terms of a light-cone parametrization does exist, and will be given in the next subsection.

\subsection{Auxiliary pole parametrization of solutions with \kap s}\label{secbehaviors} 

We begin by listing the data of the solution: we have $N+1$ poles of $\Lambda^A$ at $x_0=0$ 
and $ x_n$ with $n=1,\cdots, N$. Only one of these $N+1$ poles of $\Lambda ^A$ is double, 
while the other $N$ poles are simple. Therefore, we must have $2N$ auxiliary poles. For the solutions 
with $N$ \kap s, the functions $\l^A$ must have $N$ poles, at $x_n$. This makes sense,
because in the regular solutions, $\l^A$ was allowed to have poles only at the zeros of $\p_w H$
in the upper half-plane. But the complex zeros of our $\p_w H$ have all moved to the real axis, and so
the poles of $\l^A$ have also moved to the real axis, specifically to the points $x_n$. 

\sm

In the light-cone parametrization, one begins by considering the combination, 
\bea
L^6 = { 1 \over \l^2 - \l^6}
\eea
The $N$ poles of $\l^A$ on the real axis at $w=x_n$, 
may be realized in two different ways, namely, 
\begin{enumerate}
\itemsep -0.03 in 
 \item $L^6$ has a simple zero for $w=x_n$; or
\item $L^6$ is finite and non-zero at $w=x_n$; this behavior arises when
$\l^2$ and $\l^6$ have equal residues at $w=x_n$ so that 
 $L^A$, with $A=1,7,\dots, m+5$ have simple poles at $w=x_n$.  
\end{enumerate}
In general, a $U$-duality transformation will map one behavior into the other. 
Hence, without any loss of generality, we can construct a solution in which $L^6$ has $N$ 
simple zeros at $w=x_n$. Since $L^6(w)$ tends to a constant as $w \to \infty$
we will have,
\bea
L^6(w) = \ell ^6_\infty {S(w) \over R(w) }
\eea
where $\ell ^6 _\infty$ is a real constant, and we have introduced the following notation, 
\bea
R(w) = \prod _{p=1}^N (w-y_p)
\hskip 1in 
S(w) = \prod _{n=1}^N (w-x_n)
\eea
Since the degrees of $R$ and $S$ coincide, $L^6$ will indeed tend to a finite constant 
at $\infty$. The points $y_p\in \bR$ are the remaining $N$ auxiliary poles. Using $L^6$, 
the parametrization of (\ref{8a1}), we solve for the remaining light-cone functions. 
The zeros of $L^6$ generically become poles of $\l^A$, but the auxiliary poles of $L^6$ 
should not appear as zeros in $\l^A$. Thus, we set,
\bea
L^A(w) = \ell ^A _\infty + \sum _{p=1}^N { \ell ^A _p \over w-y_p} \label{LA-aux}
\eea
Since we must have $\Im (\l^A)=0$ for real $w$, the residues $\ell ^A_\infty$
and $\ell ^A_p$ must be real. Since $\l^2, \l^6$ should be regular at the points $y_p$, 
we make the familiar requirement (\ref{8a6}). The expression for $\ell^6 _p$ is readily 
obtained from $L^6$, and we find, 
\bea
\ell ^6 _p = \ell ^6 _\infty { S(y_p) \over R'(y_p)}
\eea
It is straightforward to check that the inequality $\bar \l \cdot \l >2$ 
is satisfied in the interior of $\Sigma$, using identically the same arguments as were used in 
\cite{Chiodaroli:2011nr} for regular solutions. This completes the explicit construction of 
solutions for the case with a single $AdS_3 \times S^3$   with $N$ \kap s.

\subsection{Calculation of the data of the covariant form}

The independent data, and their counting, of our construction are as follows,
\bea
c_0, \, \ell ^1 _\infty & \hskip 1in & 2
\\
x_n, \, y_p & & n,p=1, \cdots, N
\no \\
\ell ^R _\infty && R=6, \cdots, m+5 \hskip 1in  \hbox{totaling } m
\no \\
\ell ^R_p && R=7,8, \cdots, m+5 \hskip 0.85in \hbox{totaling } (m-1)N 
\no
\eea
giving a total of $(m+1)(N+1)+1$, of which two are $SL(2,\bR)$ artifacts. Remarkably, 
this result is the one expected from the 
counting of the physical parameters of the solutions, i.e. 
$(m+2)(N+1)$ charges obeying charge conservation and $N$ null-charge conditions 
(a total of $m+N+2$ conditions), plus the values of $m$ un-attracted scalars in the $AdS_3 \times S^3$ region.

\sm

The entanglement entropy will involve the charges $\mu_0, \mu_n$ and the 
positions $x_n$. The $\mu_n$ are readily extracted from the data, and we find,
\bea
\label{els}
\mu_n ^A  =  {\sqrt{2} c_0 R(x_n) \over \ell^6 _\infty x_n^2 S'(x_n)} L^A (x_n)
\eea
for $A=1,7, 8, \cdots, m+5$. 
The components $A=2,6$ can be readily obtained using equation (\ref{8a1}).
The expressions for $L^1(x_n)$ and $L^7(x_n) , \cdots, L^{m+5} (x_n)$ are 
provided by (\ref{LA-aux}), all other data being primary. Note that the contribution of 
$L^6(x_n)=0$ cancels out. Finally, $\mu_0$ is given in terms of $\mu_n$ by charge conservation.

\subsection{Calculation of the entanglement entropy}

Recall our key result for the entanglement entropy,
\bea
S_\ep = { \pi \over G_N} \int _{\Sigma _\ep} |dw|^2 \left ( \bar \Lambda \cdot \Lambda - 2 |\p_w H|^2 \right )
\eea
To evaluate this integral, we make use of the covariant form (\ref{cov}), as well as of the integrals (\ref{integrals}). 
The holographic 
entanglement entropy then becomes,
\bea\label{bounentncap}
S_\ep =  { \pi^2 \over G_N} \left ( \mu_0^2  \ln {1 \over \ep} + \mu_0^2 
+ \sum _{m<n} \mu_n \cdot \mu_m \ln {x_m^2 x_n^2 \over (x_m-x_n)^2} \right )   
\eea
Note that the last term contributing to the entanglement entropy  
in (\ref{bounentncap}) is a finite contribution which depends on the position 
of the auxiliary poles, i.e. on the moduli (such as the null charges) of the $N$ \kap s. 
Subtracting the UV divergent part  in (\ref{bounentncap})
 associated with the bulk contribution and using the relation (\ref{relcutoffs}),
one obtains the boundary entropy of a BCFT associated with the caps.  
This formula seems to suggest that the \kap s are providing boundary conditions 
which define a  BCFT for the bulk CFT which is fixed by the charges and value of 
scalar moduli of the asymptotic $AdS_3 \times S^3$ region. These configurations generalize 
the solution with two \kap s, discussed in  Section \ref{sec4} and depicted in Figure \ref{fig1}.

\newpage

\section{$AdS_2\times S^2$ probe-branes and the funnel solution}
\setcounter{equation}{0}
\label{sec5b}

In this section we shall construct a simple  solution which has only one asymptotic 
$AdS_3 \times S^3$ region and features one \funnel.

\subsection{The \funnel\ solution with one asymptotic $AdS_3 \times S^3$ }

\begin{figure}[htb]
\begin{center}
\includegraphics[width=6.5in]{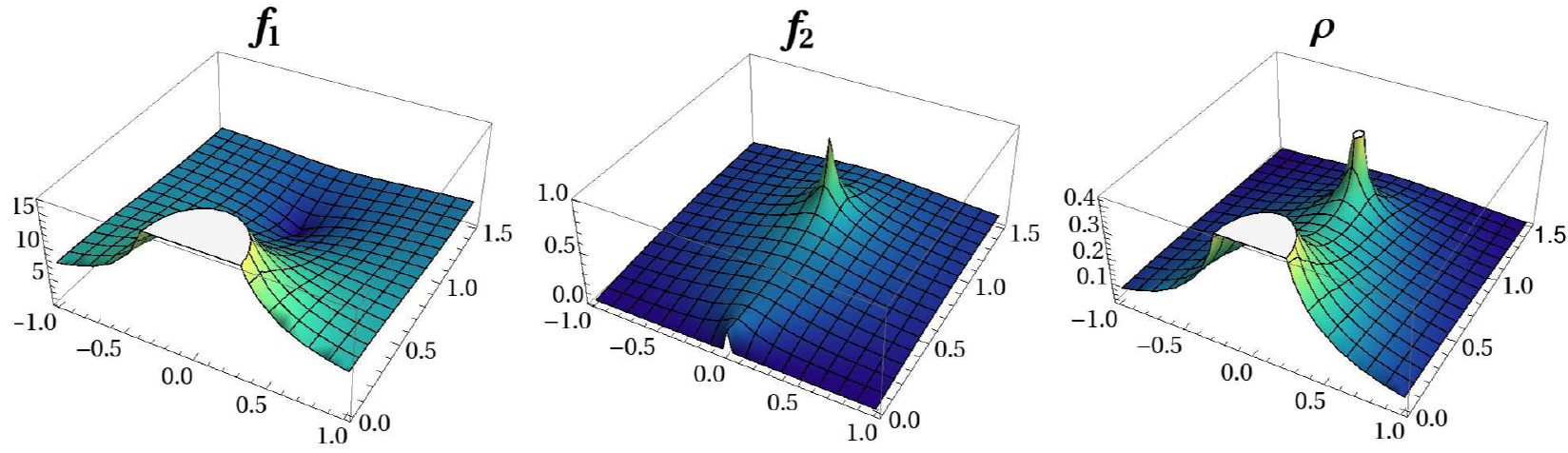} 
\caption{Plot of the metric factors for a solution with one \funnel\ at $w=i$ and
one asymptotic $AdS_3 \times S^3$ region at $w=0$. The parameters have been chosen  
randomly to have  values $\kappa = 0.74369$  and $ \mu = 0.020045 $.} 
\label{plotfunnel}
\end{center}
\end{figure}

A single asymptotic $AdS_3\times S^3$ region requires a single pole in $H$
on the real axis, which we place at $w=0$. An extra pole in the bulk of $\Sigma$
may then be chosen to be at the point $w=i$ by further use of the $SL(2,\bR)$ symmetry.
The vanishing of $\Im \lambda ^A$ on the real axis requires, however, having a mirror
pole at $w=-i$ as well. We are thus led to the following Ansatz, 
\bea
H&=& i {c_0 \over w} - i { c_0 \over \bar w}
\no \\
\Lambda^A &=&
- i {\kappa_{0}^A\over w^2} - i {\mu_{0}^A \over w} - i {\mu_{i}^A \over w - i}
- i {\mu^A _{-i}  \over w + i}  
\label{pollam}
\eea
with $c_0$ real and positive.
The associated vector of meromorphic functions $\lambda ^A$ is given by,
\bea
\lambda ^A = {w^2 \over c_0} \left (  {\kappa_{0}^A\over w^2} + {\mu_{0}^A \over w} 
+ {\mu_{i}^A \over w - i} + {\mu^A _{-i}  \over w + i}  \right ) 
\eea
Regularity as $w \to \infty$ requires overall charge conservation, 
\bea
\mu_0 ^A + \mu _i ^A + \mu _{-i} ^A=0
\eea
Reality of $\lambda ^A$ on the real axis requires $\kappa _0^A$ and $\mu_0^A$ to be real, 
as well as $\mu _{-i} ^A = \left ( \mu_i^A \right )^*$.
Combining both requirements,  we use the following parametrization, 
\bea
\mu_i ^A = - \half \mu_0 ^A + i \nu_0 ^A 
\eea
where $\nu_0 ^A$ is real. The first constraint of (\ref{1b2}) imposes 
the following conditions, 
\bea
\label{conrega}
\kappa_0 \cdot \kappa_0 &=& 2 c_0^2
\no \\
\kappa_0 \cdot  \mu_{0} = \nu_0 \cdot \mu_0 &=&0
\no \\
4 \kappa_0 \cdot \nu_0 =  4 \nu_0 \cdot \nu_0 &=& \mu_0 \cdot \mu_0
\eea
which implies $\mu_i \cdot \mu_i=0$.
The second constraint of (\ref{1b2}) imposes 
the conditions  $\mu_0^2>0$.  Similarly to the solution presented in Section \ref{sec41}, both sets of conditions 
may be solved within the restricted $SO(2,2)$ sector,  
with the help of two real parameters $\kappa >0$ and $\mu$, 
\bea
c_0 & = & \kappa / \sqrt{2}
\no \\
\kappa_0&=&\big( \kappa ,0,0,0 \big)  
\no \\
\mu_0&=& \big( 0, 2 \kappa \mu ,0,0 \big) 
\no \\
\nu_0 &=& \Big( \kappa \mu^2 , 0 , \kappa \mu \sqrt { \mu^2 -1},0 \Big)
\eea
The vector $\nu_0$ will be real, as required by our
construction, provided $\mu^2 \geq 1$. The metric factors are given by,
\bea
f_1^4 &=&   
{2 \kappa ^2 \over \mu_0^2} { \mu_0^2 \, \Im (w)^2+ \kappa ^2 |1+w^2|^2 \over |w|^4}
\no \\
f_2^4 &=&   
{2 \kappa ^2 \mu_0^2 \, \Im (w)^4 \over |w|^4\big( \mu_0^2 \, \Im (w)^2+ \kappa ^2 |1+w^2|^2 \big)}
\no \\
\rho^4&=& 
{\mu_0^2\over 8 \kappa ^2} { \mu_0^2 \, \Im (w)^2+ \kappa ^2 |1+w^2|^2 \over |w|^4 |1+w^2|^4}
\eea
Interestingly, as far as the metric factors are concerned,  
the solution has the same form as the one given for the \kap s given in Section \ref{sec41}, 
with the poles   at $w=\pm 1$ replaced by poles at $w=\pm i$.
We display plots of \funnel\ metric factors in Figure \ref{plotfunnel}.
\sm

\subsection{Calculation of the entanglement entropy}

We use (\ref{7c2}) to obtain a simple expression for the holographic 
entanglement entropy, 
\bea 
S = {  \pi \mu_0^2 \over 2 G_N} 
\int_{\Sigma_{\ep, \ep'}} { |dw|^2 \ \Im (w)^2 \over |w|^4 |w^2+1|^2} 
\eea
where we have introduced the familiar cutoff $|w|> \ep$ for the asymptotic 
$AdS_3 \times S^3$ region. The entanglement entropy for the \funnel\ clearly
also diverges at the support of the funnel, namely $w=i$. We shall use an 
independent regulator $|w-i|> \ep'$ (see Figure \ref{fig2bcft} (c)). The dependence
on both regulators is straightforward to compute, and we find,
\bea
\label{boundentfun}
S=  {\pi ^2 \mu_0^2 \over 4 G_N} 
\left ( \ln {1 \over \ep} + \ln { 1 \over \ep'}  - \half \right ) + \cO(\ep, \ep')
\eea
Comparing the entanglement entropy  (\ref{boundentfun}) with (\ref{boundenta}), 
one recognizes important similarities and differences between the solution with an 
\funnel\ and two \kap s. The first term in (\ref{boundentfun}) is the same  UV 
divergent term,  as the one in  (\ref{boundenta}) and is associated with the asymptotic 
$AdS_3 \times S^3$ region. The second term in (\ref{boundentfun}) presents a new 
divergence which is associated with the \funnel. It can be traced back to the fact that 
the location of the \funnel\ at $w=i$ is an infinite geodesic distance away from points in $\Sigma$.

\sm

Note that the holographic boundary  away from poles of $H$ is simply given by the 
boundary of the $AdS_2$ factor, which is a $0+1$-dimensional space. 
Consequently,  it seems that the \funnel\ produces a UV divergent contribution to the 
entanglement entropy from degrees of freedom which are localized on the 
$0+1$-dimensional boundary of the space where the CFT lives. 
 It would be interesting to understand the holography and the interpretation of the entanglement entropy 
for solutions with \funnel s better, but we shall leave this investigation for future work.

\newpage

\section{Discussion}
\setcounter{equation}{0}
\label{sec7}

In this paper we have relaxed the regularity conditions on the half-BPS string junction 
solutions found in \cite{Chiodaroli:2011nr}.  This allows for more general configurations 
which contain \kap s and \funnel s.  

\sm

The  \kap\   introduces a curvature singularity localized on the boundary of $\Sigma$, 
it carries a three-form charge which is null. We provided evidence that this solution  
corresponds to a fully back-reacted solution of a $D1$-brane (at its  $U$-duality 
orbit) with  $AdS_2$ world-volume.   

\sm

The  \funnel\  has no curvature, or any other, singularities, but it introduces a new 
asymptotic region emanating from the bulk of $\Sigma$ with asymptotic
$AdS_2 \times S^2 \times S^1 \times \bR^+$ geometry. We provided evidence that the 
\funnel\ corresponds to the fully back-reacted solution of a $D3$-brane (and its $U$-duality 
orbit) with $AdS_2\times S^2$ world volume, which carries electric field on  $AdS_2$.
The \funnel\ does not arise as a limiting case of the regular 
half-BPS string-junction solutions of \cite{Chiodaroli:2011nr}, in contrast to  the \kap.
The \funnel\ may be obtained as a limit, however, from solutions where $\Sigma$ has 
{\sl at least two connected boundary components}, such as those constructed 
in \cite{Chiodaroli:2009xh}. 

\sm

The simplest \kap\ and \funnel\ solutions are constructed letting the meromorphic 
functions $\l^A$ have two extra first-order poles on the boundary or in the bulk of 
$\Sigma$ respectively. In principle, we could consider, for both solutions, the limit
in which the two poles approach each other, leading to a second order pole on the boundary. 
Expanding the relevant functions close to this pole and introducing polar coordinates, 
we find that $\l \cdot \bar \l$ blows up as $\l \cdot \bar \l \simeq {\sin^2 \phi  \over r^2}$. 
Remarkably, the solution shares some of the features of the \kap\ and \funnel.
Using the expressions (\ref{7c1}) we see that both the $AdS_2$ and $S^2$ 
metric factors vanish at the pole, as expected for a \kap\ solution. 
However, the second-order pole is now at an infinite geodesic distance, 
as expected for the \funnel. 

\sm    

\begin{figure}[tb]
\begin{center}
\includegraphics[width=5.5in]{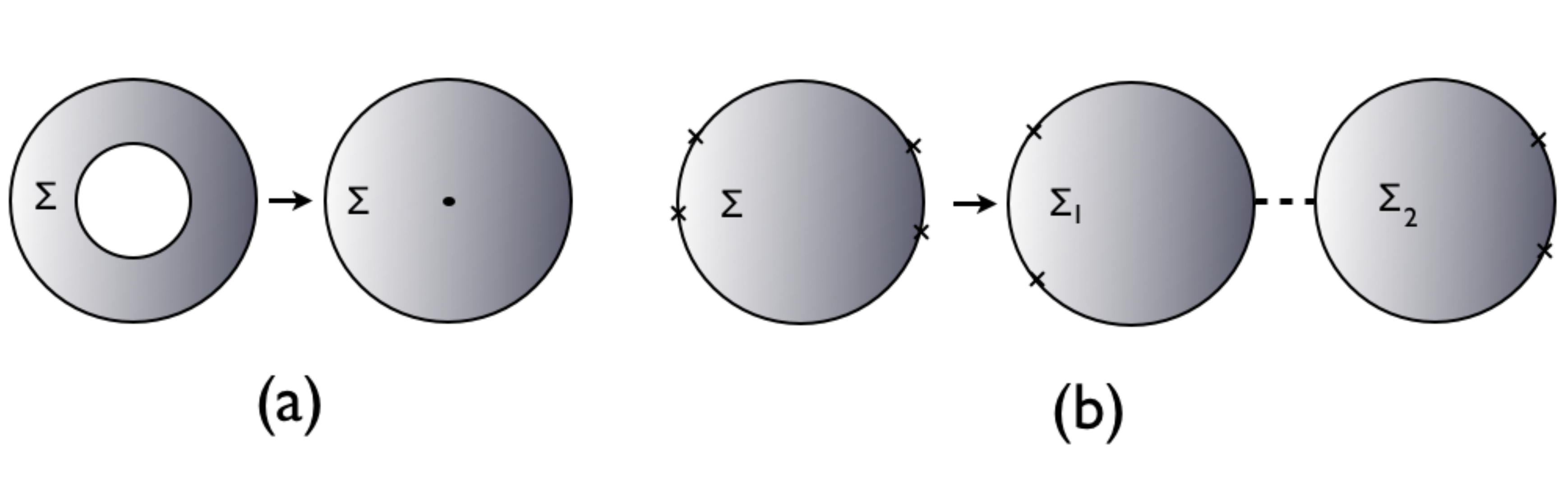} 
\caption{(a) The degeneration of an annulus produces an \funnel. 
(b) The dividing degeneration of disk produces a \kap.} 
\label{degeneration}
\end{center}
\end{figure}

The reader might ask whether these generalization are special compared to other 
singular solutions which might be constructed by general relaxation of the regularity 
conditions. The answer to this question is twofold. First, 
the \kap\ and the \funnel\ are sufficient to account for all BPS probe-branes in 
$AdS_3\times S^3$ listed in table~\ref{table2}. Second, the \kap\ and  the \funnel\ 
can be produced by taking degenerating limits of regular half-BPS junctions solutions, 
and can therefore be viewed as constituting components of the boundaries of the 
moduli space of regular solutions. We illustrate these degeneration limits in Figure \ref{degeneration}.
As was already discussed in  \cite{Chiodaroli:2009xh}, for a Riemann surface $\Sigma$ 
with more than one boundary component, the limit where one boundary degenerates to 
a point produces a new asymptotic region which we can now interpret as a funnel. 

Since the positions of the poles of $H$ on the disk $\Sigma$ are moduli of the regular 
half-BPS solution one can consider degenerations where poles come together in groups. 
For appropriately chosen charges this will lead to a dividing degeneration  with \kap s.
A more complete analysis  of the boundary of moduli space  of regular half-BPS junctions  
is under investigation \cite{MCEDMG}.

\sm

Completely regular solutions with only one asymptotic $AdS_3\times S^3$ region
are precluded by charge conservation. Thus, the simplest regular solutions are of the Janus 
type, and are dual to interface CFTs.  We made explicit use of the fact that the charge
vector of  \kap s is null  to construct solutions with only one asymptotic $AdS_3 \times S^3$ region. 
Each such solution is dual to a BCFT, and we have presented solutions with an arbitrary number
$N$  of \kap s, and discussed in detail the simplest case with $N=2$. 
Our construction turns out to be similar in spirit  to the one presented for BCFTs in four dimensions in  
\cite{Assel:2011xz,Aharony:2011yc} which, in turn,  were based on the construction of half-BPS junction 
solutions in \cite{D'Hoker:2007xy,D'Hoker:2007xz}. In \cite{Assel:2011xz,Aharony:2011yc}, the role of the  
\kap s is played by singularities associated with five-branes. In \cite{Assel:2011xz}, 
the singular  half-BPS solutions were matched precisely with the field theoretic classification of 
BCFTs \cite{Gaiotto:2008ak}. It would be interesting to explore the  BCFT dual to our  
solutions with $N$ \kap s.

\sm

Finally, we also used the  \funnel\   to construct solutions with only one asymptotic 
$AdS_3\times S^3$ region.  At present, the interpretation of these solutions remains, however,
less clear. While the holographic entanglement entropy for the BCFT, constructed 
using \kap s, gives an UV divergent part related to bulk of the CFT and a finite boundary 
entropy, the boundary entropy for the solution with the \funnel\ has an additional 
divergence coming from the funnel. One possible  interpretation is that the  
degrees of freedom residing in the funnel make an extensive contribution to the 
entanglement entropy, which is related to the additional UV-divergent term. 
In the dual CFT, a contribution this kind could not arise from a finite number of degrees of
freedom localized on the interface. It would 
be very interesting to investigate properties of the solutions containing funnels further.

\bigskip

\bigskip

\noindent{\Large \bf Acknowledgements}

\bigskip

We thank Tadashi Takayanagi for very helpful comments on a first draft of this paper.
The work of Eric D'Hoker, Michael Gutperle was supported in part by NSF grant PHY-07-57702. 
The work of Marco Chiodaroli  was supported in part by NSF grant PHY-08-55356.

\newpage

\appendix

\newpage

\section{Uplift of \kap\ solutions to ten dimensions}\label{appenda}
\setcounter{equation}{0}

To clarify the interpretation of the solutions with \kap\ singularities, 
in this appendix we will consider their uplift to ten dimensions. In general, these solutions will be characterized by 
a pole of $\lambda^A$ located on the boundary. We will fix the pole at $w=0$ throughout this section 
without any loss of generality. Then, close to the singularity we have the expansion
\be 
\l^A = {p^A \over w} + q^A + \dots 
\ee
with real $p^A$ and $q^A$. We can always find a $SO(m)$ gauge transformation rotating 
$p^A$ and $q^A$ so that the only non-zero components 
are $p^1,p^2,p^6,p^7$ and $q^1,q^2,q^6,q^7$. 
In this case, the leading behavior of the solution at the singularity will be the same as 
a $SO(2,2)$ solution which we know how to interpret in ten dimensions \cite{Chiodaroli:2009yw}. 

\sm

The translation of the old $SO(2,2)$ solutions in our language is given by the relations \cite{Chiodaroli:2011nr},
\bea 
A + \bar A &=& - 2 \sqrt{2} \big( h^1 + h^7 \big) \no \\ 
 K &=& - 2 \sqrt{2} \big( h^1 - h^7 \big) \no \\
B &=& -i \sqrt{2} L^6 
\eea
Using these relations, we can then express the ten dimensional fields of \cite{Chiodaroli:2009yw} in the auxiliary 
poles parameterization of our solutions. In particular, we will need the ten-dimensional dilaton and K3 metric factor, 
\bea 
e^{-2 \phi} &=& 2 { \Big( {(h^1)}^2 -{(h^7)}^2- {(h^6)}^2 \Big) \Big( {(h^1)}^2 -{(h^7)}^2 + ({\tilde h^6})^2 \Big) 
\over  \big( h^1 + h^7 \big)^2} 
\no \\
f_3^8 &=& 32 { \Big( {(h^1)}^2 -{(h^7)}^2- {(h^6)}^2 \Big) \Big( {(h^1)}^2 -{(h^7)}^2 + ({\tilde h^6})^2 \Big) 
\over  \big( h^1 - h^7 \big)^2}  
\eea
As explained in Section \ref{secbehaviors}, 
a simple pole of $\l^A$ on 
the boundary corresponds to two possible behaviors of the functions $L^A$:
either  $L^6$ has a simple zero while the functions $L^A$ ($A=1,7,\dots,m+5$) 
are non-zero, or $L^6$ is non-zero, but the functions $L^A$ have first order poles. 
In general there exists a $U$-duality transformation mapping the two different behaviors into each other. 
Without any loss of generality we will study the second kind of behavior expanding the relevant functions
 as follows,
\bea 
L^6 &=& a^6_{(0)} + a^6_{(1)} w +  \dots \no \\
 L^A &=& {\ell^A_0  \over w }+  a^A_{(0)}  + a^A_{(1)} w + \dots \ , \qquad A=1,7  \no \\
H &=& i h_{(1)} w + c.c.   
\eea 
with ${\ell^1}^2 = {\ell^7}^2$. 
There are two possibilities. If $\ell_0^7=  + \ell^1_0 $, after some algebra we get,
\bea 
e^{-2 \phi} &=& 4 {\ell^1_0 \Big( ({a_{(0)}^6})^2 + 2 \ell_0^1 \big( a_{(1)}^7 - a_{(1)}^1 \big)  \sin^2 \phi  
\Big) \over \big( a_{(1)}^7 - a_{(1)}^1 \big) r^2 } 
\no \\ 
f_3^8 &=& 16 {  \big( a_{(1)}^7 - a_{(1)}^1 \big) \Big( ({a_{(0)}^6})^2 + 2 \ell_0^1 \big( a_{(1)}^7 - a_{(1)}^1 \big)  \sin^2 \phi  
\Big) \over \ell^1_0 } r^2 
\eea 
In this case we need $a_{(1)}^7 - a_{(1)}^1 >0$ for the solution to satisfy the inequality $\l \cdot \bar \l > 2$.  
The radial behavior is exactly the one expected from a fundamental $D5$-brane. 
Similarly, in case $\ell_0^7= - \ell^1_0 $ we obtain 
the expressions,
\bea 
e^{-2 \phi} &=& { - \big( a_{(1)}^1 + a_{(1)}^7 \big) \Big( ({a_{(0)}^6})^2 - 2 
\ell_0^1 \big( a_{(1)}^1 + a_{(1)}^7 \big)  \sin^2 \phi  
\Big) \over \ell^1_0   } r^2  \\ 
f_3^8 &=& 64 { \ell^1_0  \Big( ({a_{(0)}^6})^2 + 2 \ell_0^1 \big( a_{(1)}^7 - a_{(1)}^1 \big)  \sin^2 \phi  
\Big) \over  - \big( a_{(1)}^1 + a_{(1)}^7 \big) r^2  } 
\eea 
with $ a_{(1)}^7 + a_{(1)}^1 < 0$. This $r$-dependence is the one expected for a fundamental $D1$-brane. 
All other \kap\ singularities, e.g. the one corresponding to fundamental strings and $NS5$-branes, 
can be generated acting with a $U$-duality transformation on the solutions above.

\section{Relating the AdS and CFT UV-regulators}\label{appendb}

In this appendix we discuss the relation of the UV cutoff in the bulk AdS and the UV cutoff 
on the CFT side. The discussion follows  \cite{Azeyanagi:2007qj,Chiodaroli:2010ur} 
and is included here  to make the present paper self-contained.

\sm

Near a pole of $H$ the metric becomes asymptotically $AdS_3\times S^3$, since the three sphere is not important for the relation of the cutoffs we will only consider the three dimensional part of the metric which takes the following form for  $x\to \infty$ and $z$ finite.
\bea\label{boundlima}
\lim_{x\to  \infty} ds^2= R_{AdS_3}^2\left(  dx^2 +\lambda   e^{2x} {dz^2-dt^2\over z^2}\right) + o(e^{-2x})
\eea
One can absorb the constant $\lambda $ by a shift in $x$,
\be\label{shiftx}
x=\tilde x  - {1\over 2} \ln (\lambda )
\ee
and one gets
\bea\label{asymmetb}
\lim_{x\to  \infty} ds^2= R_{AdS_3}^2\left(  d\tilde x^2 +  e^{2\tilde x} {dz^2-dt^2\over z^2}\right) + o(e^{-2\tilde x})
\eea
In this limit we can perform an coordinate change which maps the $AdS_2$ slicing to a Poincare slicing by
\bea\label{fgchangeb}
\tilde x\to +\infty, \quad \xi\to 0, \;\eta>0 &:& \quad \quad e^{-2\tilde x} =  {\xi^2 \over \eta^2}, \quad  z= \eta \Big(1+ {1\over 2} {\xi^2\over \eta^2} \Big)
\eea
The  terms as $\xi \to 0$ of the metric becomes
\be\label{adsmetp}
\lim_{|x| \to \pm \infty} ds^2= {R_{AdS_3}^2\over \xi^2} \Big( d\xi^2+ d \eta^2 -dt^2\Big)+ o(1)
\ee
The  half-space on which the BCFT lives has  $\eta>0$. 
Note that near the boundary $\eta=0$  the change of coordinates is more complicated.
In particular, the coordinate change (\ref{fgchangeb}) is not  smooth at $\eta=0$.  
For the purposes of calculating the holographic entanglement entropy  we consider 
a point  $|\eta|=z_0>0$ and never approach $\eta=0$.  However, for other calculations, 
such as  the    boundary OPE, the limit $\eta\to 0$ has to be considered and the 
discussion is much more involved.

The  metric (\ref{adsmetp}) in the limit $x\to \infty, \xi\to 0$ takes the standard form of 
$AdS_3$ in Poincar\'e coordinates and  standard rules of AdS/CFT apply in identifying a   
cutoff $\xi_{UV}$  for the coordinate $\xi$ and the UV cutoff of the CFT. It follows from 
(\ref{fgchangeb}) and (\ref{shiftx}) that the UV cutoff of the CFT and the cutoff in the 
$AdS$-slicing coordinate $x$ are related by
\bea
\label{cutoffa}
x\to +\infty&:& \quad \xi_{UV} = {2 z_0 \over \sqrt{\lambda_+} }e^{-x_\infty}
\eea

\section{Calculating entanglement entropy for an \funnel}

The holographic entanglement entropy for the funnel is given by, 
\bea 
S = {  \pi \mu_0^2 \over 2 G_N} I_f
\hskip 1in 
J_c =  \int_{\Sigma_{\ep, \ep'}} { |dw|^2 \ \Im (w)^2 \over |w|^4 |w^2+1|^2} 
\eea
where we have introduced the familiar UV cutoff $|w|> \ep$ for the asymptotic 
$AdS_3 \times S^3$ region. The entanglement entropy for the \funnel\ clearly
also diverges at the points of support of the funnel, namely $w=i$. This is an IR divergence,
for which we use an independent regularization $|w-i|> \ep'$. To evaluate the integrals, 
we set $w=y+ix$, so that,
\bea
I_f =  \int _{-\infty} ^\infty dy \int _0 ^\infty dx 
{ x^2 \theta _\ep (x,y) \theta _{\ep'} (x-1,y) \over (x^2+y^2)^2 ((x-1)^2 +y^2) ((x+1)^2+y^2)}
\eea
We split up this integral as follows, 
\bea
I_f = -I_c + I_f ^{y>1} + I_f ^{y<1} + I_f ^{\ep, \ep'} 
\eea
where
\bea
I_f ^{y>1} & = & 
\int _{-\infty}  ^\infty dx \int _1 ^\infty dy 
{1  \over (x^2+y^2) ((x-1)^2 +y^2) ((x+1)^2+y^2)}
\no \\
I_f ^{y<1} & = & 
\int _{-\infty}  ^\infty dx \int _0 ^1 dy \left ( 
{ 1  \over (x^2+y^2) ((x-1)^2 +y^2) ((x+1)^2+y^2)} - \sum _{t=0,\pm1} {(1+|t|)^{-2} \over ((x-t)^2+y^2)} \right )
\no \\
I^{\ep, \ep'} _f & = & \int _{-\infty}  ^\infty dx \int _0 ^1 dy
{ \theta _{\ep} (x,y) \over (x^2+y^2) } + {1 \over 4} \sum _{t=\pm 1}  \int _{-\infty}  ^\infty dx \int _0 ^1 dy
{ \theta _{\ep'} (x-t,y)  \over ((x-t)^2+y^2) }
\eea
Evaluating these integrals  gives,
\bea
I_f ^{y>1} & = & \pi \ln 5 - {9 \pi \over 4} \ln 2
\no \\
I_f ^{y<1} & = & - \pi \ln 5 +{\pi \over 4} \ln 2
\no \\
I_f ^{\ep, \ep'} & = & \pi \ln 2 + {\pi \over 2} \ln 2 + \pi \ln {1 \over \ep} + {\pi \over 2} \ln { 1 \over \ep'}
\eea
$I_c$ is the integral encountered in the calculation for the \kap, 
\be I_c =  \int_{\Sigma_\ep} { |dw|^2 \ \Im (w)^2 \over |w|^4 |1-w^2|^2} =  {\pi \over 2}  \ln {1 \over \ep}  - {\pi \over 2} \ln 2 + {\pi \over 4} \ee

Putting all together, we find, 
\bea
I_f = {\pi \over 2} \ln {1 \over \ep} + {\pi \over 2} \ln { 1 \over \ep'} -  {\pi \over 4}
\eea
which gives the following expression for the funnel entropy,
\bea
S =  { \pi^2 \mu_0^2 \over 4 G_N} \left ( \ln {1 \over \ep} + \ln {1 \over \ep'}  - \half \right )   
\eea

\end{document}